\documentclass[nofootinbib,aps,prd,superscriptaddress,preprintnumbers,%
 showpacs,showkeys,floatfix,amssymb,amsfonts]{revtex4-1}  

\usepackage{graphicx}
\usepackage{bm}
\usepackage{amsmath}
\usepackage{amssymb}
\usepackage{amsfonts}
\usepackage{float}
\usepackage{dsfont}  
\usepackage{slashed}  
\usepackage{booktabs}
\usepackage{multirow}
\usepackage{subfigure}
\usepackage[sort&compress]{natbib}
\usepackage{ulem}
\usepackage{empheq}
\usepackage{color}
\usepackage{tipa}

\newcommand{\be}{\begin{equation}}  
\newcommand{\ee}{\end{equation}}  
\newcommand{\beq}{\begin{eqnarray}} 
\newcommand{\eeq}{\end{eqnarray}}

\newcommand{\bea}{\begin{eqnarray}}
\newcommand{\eea}{\end{eqnarray}}

\newcommand{\MSb}{{\overline{\rm MS}}}

\usepackage{xr-hyper}

\usepackage[svgnames,x11names,table]{xcolor}
\usepackage[colorlinks=true,linkcolor=MediumBlue,citecolor=MediumBlue,urlcolor=MediumBlue]{hyperref}
\usepackage{hyperref}

\begin{document}
\title{Generalized Parton Distributions from Lattice QCD \\[1ex] with Asymmetric Momentum Transfer: Axial-vector case}
\author{Shohini Bhattacharya}
\email{sbhattach@bnl.gov}
\affiliation{RIKEN BNL Research Center, Brookhaven National Laboratory, Upton, NY 11973, USA}
\author{Krzysztof Cichy}
\author{Martha Constantinou}
\email{marthac@temple.edu}
\affiliation{Department of Physics,  Temple University,  Philadelphia,  PA 19122 - 1801,  USA}
\author{Jack Dodson}
\affiliation{Department of Physics,  Temple University,  Philadelphia,  PA 19122 - 1801,  USA}
\author{Xiang Gao}
\affiliation{Physics Division, Argonne National Laboratory, Lemont, IL 60439, USA}
\author{Andreas Metz}
\affiliation{Department of Physics,  Temple University,  Philadelphia,  PA 19122 - 1801,  USA}
\author{Joshua Miller}
\email{joshua.miller0007@temple.edu}
\affiliation{Department of Physics,  Temple University,  Philadelphia,  PA 19122 - 1801,  USA}
\author{Swagato Mukherjee}
\affiliation{Physics Department, Brookhaven National Laboratory, Upton, New York 11973, USA}
\author{Peter Petreczky}
\affiliation{Physics Department, Brookhaven National Laboratory, Upton, New York 11973, USA}
\author{Fernanda Steffens}
\affiliation{Institut f\"ur Strahlen- und Kernphysik, Rheinische Friedrich-Wilhelms-Universit\"at Bonn,\\ Nussallee 14-16, 53115 Bonn}
\author{Yong Zhao}
\affiliation{Physics Division, Argonne National Laboratory, Lemont, IL 60439, USA}
%
\begin{abstract}
Recently, we made significant advancements in improving the computational efficiency of lattice QCD calculations for Generalized Parton Distributions (GPDs). 
This progress was achieved by adopting calculations of matrix elements in asymmetric frames, deviating from the computationally-expensive symmetric frame typically used, and allowing freedom in the choice for the distribution of the momentum transfer between the initial and final states. 
A crucial aspect of this approach involves the adoption of a Lorentz covariant parameterization for the matrix elements, introducing Lorentz-invariant amplitudes. 
This approach also allows us to propose an alternative definition of quasi-GPDs, ensuring frame independence and potentially reduce power corrections in matching to light-cone GPDs. 
In our previous work, we presented lattice QCD results for twist-2 unpolarized GPDs ($H$ and $E$) of quarks obtained from calculations performed in asymmetric frames at zero skewness. Building upon this work, we now introduce a novel Lorentz covariant parameterization for the axial-vector matrix elements. We employ this parameterization to compute the axial-vector GPD $\widetilde{H}$ at zero skewness, using an $N_f=2+1+1$ ensemble of twisted mass fermions with clover improvement. The light-quark masses employed in our calculations correspond to a pion mass of approximately 260 MeV. 
\end{abstract}
\maketitle

\section{Introduction}
\label{s:intro}
Parton distribution functions (PDFs) play a crucial role in understanding the quark and gluon structure of strongly interacting systems~\cite{Collins:1981uw}. 
These functions, measurable in processes such as inclusive deep-inelastic lepton-nucleon scattering, provide valuable insights into the distribution of partons within hadrons as a function of their momentum fraction, denoted as $x$. 
PDFs are defined through matrix elements of bi-local operators, where the parton fields are separated by a light-like interval, and the operators are evaluated for the same initial and final hadron states. 
Generalized parton distributions (GPDs) extend the concept of PDFs by considering light-like parton operators computed for different initial and final states~\cite{Mueller:1998fv, Ji:1996ek, Radyushkin:1996nd}. 
GPDs introduce additional dependencies on the longitudinal momentum transfer ($\xi$) and the invariant momentum transfer ($t$) to the target, in addition to the parton momentum fraction ($x$). 
While this multi-variable nature makes GPDs more complex, they offer a wealth of information beyond PDFs. 
In particular, GPDs provide three-dimensional images of hadrons~\cite{Burkardt:2000za, Ralston:2001xs, Diehl:2002he, Burkardt:2002hr}, enable access to the angular momenta of partons~\cite{Ji:1996ek}, and offer insights into the pressure and shear forces within hadrons~\cite{Polyakov:2002wz, Polyakov:2002yz, Polyakov:2018zvc}. 
Recently, it has been discovered that GPDs exhibit chiral and trace anomaly poles, which provide insights into phenomena such as mass generation in QCD, chiral symmetry breaking, and confinement~\cite{Bhattacharya:2022xxw,Bhattacharya:2023wvy,Bhattacharya:2023ksc}. 
Understanding these imprints can offer valuable insights into fundamental aspects of QCD. 
We also refer the reader to several other review articles that extensively discuss the physics of GPDs~\cite{Goeke:2001tz, Diehl:2003ny, Ji:2004gf, Belitsky:2005qn, Boffi:2007yc, Guidal:2013rya, Mueller:2014hsa, Kumericki:2016ehc}.

Experimental knowledge about GPDs can be acquired through hard exclusive scattering processes such as deep virtual Compton scattering~\cite{Mueller:1998fv, Ji:1996ek, Radyushkin:1996nd, Ji:1996nm, Collins:1998be} and hard exclusive meson production~\cite{Radyushkin:1996ru, Collins:1996fb, Mankiewicz:1997uy}. 
However, extracting GPDs from these reactions in a model-independent manner is highly intricate, primarily due to the integration over the momentum fraction $x$ in observable quantities such as Compton form factors. 
A recent discussion and detailed analysis of this issue can be found in Ref.~\cite{Bertone:2021yyz}. 
Efforts to parameterize the GPDs and fit them from global experiments can be found in Refs.~\cite{Polyakov:2002wz,Guidal:2004nd,Goloskokov:2005sd,Mueller:2005ed,Kumericki:2009uq,Goldstein:2010gu,Gonzalez-hernandez:2012xap,Kriesten:2021sqc,Hashamipour:2021kes,Guo:2022upw,Guo:2023ahv,Hashamipour:2022noy}, as well as the impact of MINERvA measurement of the antineutrino-proton scattering cross section on
the axial-vector GPDs at zero skewness in Ref.~\cite{Irani:2023lol}. 
However, the majority of the studies are still in the initial stages.
Therefore, obtaining information on GPDs directly from first principles in lattice QCD is highly desirable. 
However, direct calculations of GPDs on a Euclidean lattice are prohibited due to their light-cone definition. 
Limited information on GPDs has been traditionally obtained from their Mellin moments (see, e.g., Refs.~\cite{Hagler:2003jd, QCDSF-UKQCD:2007gdl, Alexandrou:2011nr, Alexandrou:2013joa,Constantinou:2014tga}), with simulations at the physical point becoming available in recent years~\cite{Green:2014xba,Alexandrou:2017ypw,Alexandrou:2017hac,Hasan:2017wwt,Gupta:2017dwj,Capitani:2017qpc,Alexandrou:2018sjm,Shintani:2018ozy,Bali:2018qus,Bali:2018zgl,Alexandrou:2019ali,Jang:2019jkn,Constantinou:2020hdm,Alexandrou:2022dtc,Jang:2023zts}. 
Despite the progress made, the precise dependence on $x$ has remained elusive.

Over the last few years, the emergence of alternative approaches for accessing GPDs in momentum space has sparked a highly promising research program in lattice QCD. 
In our work, we use the quasi-distributions method~\cite{Ji:2013dva}, which entails calculating matrix elements involving momentum-boosted hadrons and non-local operators. 
To establish a connection with light-cone GPDs, we employ the framework of Large-Momentum Effective Theory (LaMET)~\cite{Ji:2014gla,Ji:2020ect}. 
Extensive reviews, including other methods for obtaining $x$-dependent distribution functions, can be found in Refs.~\cite{Cichy:2018mum,Ji:2020ect,Constantinou:2020pek,Cichy:2021lih,Cichy:2021ewm}. 
While calculations of PDFs, the simplest one-dimensional distributions, dominate the literature, applications concerning GPDs, three-dimensional distributions with novel aspects, remain relatively limited (see e.g., Refs.~\cite{Ji:2015qla, Xiong:2015nua, Bhattacharya:2018zxi, Liu:2019urm, Bhattacharya:2019cme, Chen:2019lcm, Radyushkin:2019owq, Ma:2019agv, Luo:2020yqj, Alexandrou:2020zbe,Alexandrou:2021bbo,CSSMQCDSFUKQCD:2021lkf,Dodson:2021rdq, Ma:2022ggj, Shastry:2022obb, Ma:2022gty,Yao:2022vtp}).

In our recent publication of Ref.~\cite{Bhattacharya:2022aob}\footnote{See also Refs.~\cite{Bhattacharya:2023tik,Constantinou:2022fqt,Cichy:2023dgk}.}, we achieved significant advancements in enhancing the computational efficiency of lattice QCD calculations for off-forward matrix elements. The work was also extended to calculate the Mellin moments of the unpolarized GPDs~\cite{Bhattacharya:2023ays}, including high moments enabling a physical picture of quark distribution in the transverse plane. In these calculations, we employed a unique approach using asymmetric frames, which differ from the more commonly used symmetric frames. In this approach, the entire momentum transfer $\Delta$ is applied to the initial state (source) of the nucleon. This choice not only reduces the computational cost but also offers the advantage of covering a broader range in $t \equiv \Delta^2$, enabling us to effectively map the GPDs across a larger $t$-space. In our previous work, our focus was on unpolarized quark GPDs ($H$ and $E$) at zero skewness. We introduced a novel Lorentz-covariant parameterization for the vector matrix element in terms of Lorentz-invariant amplitudes. 
This also allowed us to establish connections between matrix elements from any two kinematic frames. 
Additionally, we employed this amplitude-based approach to propose a frame-independent definition of quasi-GPDs and demonstrated that these definitions can potentially result in reduced power corrections in the matching relations to light-cone GPDs. 
In this work, we extend the amplitude-based approach to compute the axial-vector GPD $\widetilde{H}$ at zero skewness. 
For a comprehensive discussion on the inaccessibility of $\widetilde{E}$ at $\xi=0$, we refer to Sec.~\ref{s:para}, where we discuss the intricacies and reasons behind this limitation.

The paper is structured as follows. 
In Section II, we begin by presenting the definitions of axial-vector light-cone and quasi-GPDs. 
We then shift our focus towards discussing the Lorentz-covariant decomposition of axial-vector matrix elements in terms of the Lorentz-invariant amplitudes. 
Furthermore, we establish the relations between these amplitudes and the GPDs $\widetilde{H}$ and $\widetilde{E}$. 
Based on these amplitudes, we propose a few potential candidates for a new, frame-independent definition of quasi-GPDs under the constraints of finite boost momentum. 
We thoroughly explore the interpretations of these new definitions, carefully examining the subtleties involved while also addressing the important issue of uniqueness/non-uniqueness in their formulation. 
In Section III, we provide the Euclidean decompositions of lattice-calculable matrix elements in terms of these amplitudes. 
We also outline our lattice setup for the calculations in position space. 
Section IV is dedicated to our numerical results, accompanied by a detailed comparison between the symmetric and asymmetric frames at different stages, both in coordinate space and momentum space. 
Notably, we provide numerical results for the invariant amplitudes and the twist-2 light-cone GPD $\widetilde{H}$, specifically for $\xi=0$. 
Finally, in Section V, we conclude our findings and discuss potential future prospects for further research and exploration in this field.

\section{Strategy of frame transformation}
\label{s:para}

Computing GPDs in the symmetric frame presents significant challenges in lattice QCD. 
Extracting a range of momentum transfers requires separate calculations for each $\Delta$, severely limiting the accessible momentum transfer range. 
This prompts the question of calculating GPDs in computationally advantageous asymmetric frames. 
One approach, as outlined in our previous work~\cite{Bhattacharya:2022aob}, establishes a connection between the symmetric and asymmetric frames through a suitable Lorentz transformation. 
Employing a Lorentz transformation along the $z$-direction does not work since a spatial operator distance will receive a nonzero temporal component, which cannot be dealt with in lattice-QCD calculations. In contrast, transverse Lorentz transformations (``transverse boosts") preserve the spatial operator distance.
In our second approach, we have developed a Lorentz covariant formalism that allows calculations in any frame. 
By parameterizing the relevant matrix element using Lorentz-invariant (frame-independent) amplitudes, we establish connections between different frames. 
In the following sections, we will explore this approach and its implications for computing axial-vector GPDs in asymmetric frames.

\subsection{Definitions of GPDs}
\label{s:def_GPDs}
To begin, let us revisit the definition of light-cone quark GPDs for a spin-1/2 hadron. 
In position space, GPDs characterize non-local quark field matrix elements, which are defined as follows:
\begin{equation}
F^{[\Gamma]}(z^-, \Delta, P)= \langle p_f,\lambda'| \bar{\psi} (-\tfrac{z}{2})\, \Gamma \, {\cal W}(-\tfrac{z}{2},\tfrac{z}{2}) \psi (\tfrac{z}{2})|p_i,\lambda \rangle \Big |_{z^{+}=0,\vec{z}_{\perp}=\vec{0}_{\perp}} \, ,
\label{e:corr_standard_GPD}
\end{equation} 
where $\Gamma$ represents a gamma matrix. 
The gauge invariance of this correlator is ensured by the presence of the Wilson line
\begin{equation}
{\cal W}(-\tfrac{z}{2},\tfrac{z}{2})\Big|_{z^{+}=0,\vec{z}_{\perp}=\vec{0}_{\perp}} = {\cal P} \, {\rm exp} \bigg ( -ig \int^{\tfrac{z^{-}}{2}}_{-\tfrac{z^{-}}{2}} \, dy^{-} A^{+}(0^{+},y^{-},\vec{0}_{\perp}) \bigg ) \,. 
\label{e: wilson_line_standard_GPD}
\end{equation}
In Eq.~(\ref{e: wilson_line_standard_GPD}), the parameter $g$ represents the strong coupling constant, while $A^{+}$ denotes the plus-component of the gluon field on the light cone. 
The initial (final) hadronic state in Eq.~(\ref{e:corr_standard_GPD}) is described by its 4-momentum $p_i$ ($p_f$) and helicity $\lambda$ ($\lambda'$). 
We introduce the following kinematic variables: the average 4-momentum of the hadrons $P$, the (aforementioned) 4-momentum transfer $\Delta$, the skewness $\xi$ (which is defined for hadrons with a large light-cone plus-momentum and represents the longitudinal momentum transfer to the hadron), and the (aforementioned) invariant squared 4-momentum transfer $t$,
\begin{equation}
P= \frac{1}{2}(p_i+p_f), \qquad
\Delta =p_f-p_i, \qquad 
\xi =\frac{p^{+}_i-p^{+}_f}{p^{+}_i+p^{+}_f}, \qquad
t = \Delta^{2} \, .
\label{e:kinematics}
\end{equation}
We use the definitions in Eq.~\eqref{e:kinematics} in both the symmetric and asymmetric frames.
At twist-2, the correlator with $\Gamma = \gamma^+ \gamma_5$ in Eq.~(\ref{e:corr_standard_GPD}) can be characterized by two distinct axial-vector GPDs, $\widetilde{H}$ and $\widetilde{E}$. 
In position space, the expression is given by~\cite{Diehl:2003ny}
\begin{align}
F^{[\gamma^+ \gamma_5]} (z^-, \Delta, P)&  =  \bar{u}(p_f, \lambda ') \bigg [ \gamma^+ \gamma_5 \widetilde{H} (z^-, \xi, t) +  \frac{\Delta^+ \gamma_5}{2m} \widetilde{E} (z^-, \xi, t) \bigg ] u(p_i, \lambda) \, .
\label{e:GPD_def_pos}
\end{align}
For the expression corresponding to Eq.~(\ref{e:corr_standard_GPD}) in momentum space, the Fourier transform is taken with respect to $P \cdot z$ while keeping $P^+$ fixed, leading to
\begin{align}
F^{[\gamma^+ \gamma_5]} (x, \Delta)&  =  \dfrac{1}{2P^+} \int \dfrac{d(P \cdot z)}{2\pi} e^{i x P \cdot z}  F^{[\gamma^+ \gamma_5]} (z^-, \Delta, P) \nonumber \\
& = \dfrac{1}{2P^+} \bar{u}(p_f, \lambda ') \bigg [ \gamma^+ \gamma_5 \widetilde{H} (x, \xi, t) +  \frac{\Delta^+ \gamma_5}{2m} \widetilde{E} (x, \xi, t) \bigg ] u(p_i, \lambda) \, .
\label{e:GPD_def}
\end{align}

Now, let us redirect our focus to quasi-GPDs, which are defined in position space through the equal-time correlator~\cite{Ji:2013dva}
\begin{equation}
F^{[\Gamma]}(z^3, \Delta, P) =  \langle p_f,\lambda '| \bar{\psi}(-\tfrac{z}{2}) \, \Gamma \, {\cal W}(-\tfrac{z}{2}, \tfrac{z}{2}) \psi (\tfrac{z}{2})|p_i, \lambda \rangle \Big |_{z^{0}=0, \vec{z}_{\perp}=\vec{0}_{\perp}} \, .
\label{e: corr_quasi_GPD_1}
\end{equation}
Here, the Wilson line is given by 
\begin{equation}
{\cal W}(-\tfrac{z}{2},\tfrac{z}{2})\Big|_{z^{0}=0,\vec{z}_{\perp}=\vec{0}_{\perp}} = {\cal P} \, {\rm exp} \bigg ( -ig \int^{\tfrac{z^{3}}{2}}_{-\tfrac{z^{3}}{2}} \, dy^{3} A^{3}(0, \vec{0}_{\perp}, y^{3}) \bigg )\, .
\end{equation}
For $\Gamma = \gamma^3 \gamma_5$, one finds
\begin{align}
F^{[\gamma^3 \gamma_5]} (z^3, \Delta, P) & = \bar{u}(p_f, \lambda ') \bigg [ \gamma^3 \gamma_5 \widetilde{\mathcal{H}}_3 (z^3, \xi, t; P^3) +  \frac{\Delta^3 \gamma_5}{2m} \widetilde{\mathcal{E}}_3 (z^3, \xi, t;P^3) \bigg ] u (p_i, \lambda) \, ,
\label{e:qGPD_def_pos}
\end{align}
with the quasi-GPDs $\widetilde{\mathcal{H}}_3 (z^3, \xi, t; P^3)$ and $\widetilde{\mathcal{E}}_3 (z^3, \xi, t; P^3)$. Eq.~\eqref{e:qGPD_def_pos} is the quasi-GPD counterpart of Eq.~\eqref{e:GPD_def_pos}.
For the expression corresponding to Eq.~(\ref{e:qGPD_def_pos}) in momentum space, we perform a Fourier transform with respect to $P \cdot z$ while keeping $P^3$ fixed, yielding  
\begin{align}
F^{[\gamma^3 \gamma_5]}(x, \Delta; P^3) & =  \dfrac{1}{2P^0}\int \frac{d(P \cdot z)}{2\pi}  e^{i x P \cdot z} F^{[\gamma^3 \gamma_5]} (z^3, \Delta, P) \nonumber \\
& = \dfrac{1}{2P^0} \bar{u}(p_f, \lambda ') \bigg [ \gamma^3 \gamma_5 \widetilde{\mathcal{H}}_3 (x, \xi, t; P^3) +  \frac{\Delta^3 \gamma_5}{2m} \widetilde{\mathcal{E}}_3 (x, \xi, t;P^3) \bigg ] u (p_i, \lambda) \, .
\label{e: corr_quasi_GPD}
\end{align}
In Ref.~\cite{Constantinou:2017sej}, it was proposed to define the quasi-counterpart of the light-cone GPDs $\widetilde{H}$ and $\widetilde{E}$ using $\Gamma = \gamma^3 \gamma_5$, as presented in Eq.~\eqref{e:qGPD_def_pos}. 
The rationale for selecting $\gamma^{3}\gamma_{5}$ instead of, for instance, $\gamma^0 \gamma_5$ is the absence of mixing with other operators under renormalization, where this mixing is regarded as a lattice artifact caused by chiral symmetry breaking~\cite{Constantinou:2017sej}. 
Furthermore, Ref.~\cite{Bhattacharya:2019cme} argues that, for this definition, it becomes necessary to substitute $\gamma^+ \gamma_5/P^+$ with $\gamma^3\gamma_5/P^0$ in the prefactor of $\widetilde{H}$ to ensure consistency with the forward limit.
The definition in Eq.~(\ref{e: corr_quasi_GPD}) also produces the correct local limit when integrated with respect to $x$.

\subsection{Parameterization of an axial-vector matrix element}
\label{s:decomposition}
Now, let us discuss the Lorentz-covariant decomposition of the axial-vector matrix elements, specifically Eq.~\eqref{e:corr_standard_GPD} with $\Gamma = \gamma^\mu \gamma_5$, for spin-1/2 particles in position space.
By incorporating parity constraints, we establish that the axial-vector matrix element can be expressed as a combination of eight distinct Dirac structures, each multiplied by a corresponding Lorentz-invariant amplitude. 
The choice of basis for the amplitudes is not unique, and here we employ 
\begin{align}
\widetilde{F}^{\mu} (z, P, \Delta)
& \equiv \langle p_f;\lambda'| \bar{\psi} (-\tfrac{z}{2})\, \gamma^\mu \gamma_5 \, {\cal W}(-\tfrac{z}{2},\tfrac{z}{2}) \psi (\tfrac{z}{2})|p_i;\lambda \rangle \nonumber \\[0.3cm]
& = \bar{u}(p_f,\lambda') \bigg [ \dfrac{i \epsilon^{\mu P z \Delta}}{m} \widetilde{A}_1 + \gamma^{\mu} \gamma_5 \widetilde{A}_2 + \gamma_5 \bigg ( \dfrac{P^\mu}{m} \widetilde{A}_3 + m z^\mu \widetilde{A}_4 + \dfrac{\Delta^\mu}{m} \widetilde{A}_5 \bigg ) \nonumber \\[0.1cm]
& \hspace{1.65cm} + m \slashed{z}\gamma_5 \bigg ( \dfrac{P^\mu}{m} \widetilde{A}_6 + m z^\mu \widetilde{A}_7 + \dfrac{\Delta^\mu}{m} \widetilde{A}_8 \bigg )\bigg ] u(p_i, \lambda) \, ,
\label{helicity_para}
\end{align}
where $\epsilon^{\mu P z \Delta}=\epsilon^{\mu \alpha \beta \gamma} P_\alpha z_\beta \Delta_\gamma$. We note that the above equation holds for a general value of $z$ and has a smooth $z\to0$ limit.
The amplitudes $\widetilde{A}_i$ are frame-independent, while the basis vectors are generally frame-dependent.
Note also that the basis vectors in Eq.~\eqref{helicity_para} do not contain factors of $z^2$.
(For example, such factors can occur when working with an orthogonal set of basis vectors.)
For the amplitudes, we adopt the concise notation $\widetilde{A}_i \equiv \widetilde{A}_i (z\cdot P, z \cdot \Delta, \Delta^2, z^2)$ for brevity.
The procedure of deriving these results closely follows the steps outlined in Ref.~\cite{Meissner:2009ww}, with a similar treatment found in Ref.~\cite{Rajan:2017cpx} where the matrix element was parameterized in momentum space using a straight Wilson line. 
It is worth noting that the number of amplitudes is the same as for the vector current ($\Gamma = \gamma^\mu$) discussed in Ref.~\cite{Bhattacharya:2022aob}. 
While it is possible to work with alternative sets of basis vectors, the number of independent amplitudes will remain unchanged, requiring eight independent lattice matrix elements to disentangle all the amplitudes. 
Furthermore, in Appendix~\ref{s:symmetries}, we present a comprehensive discussion on the symmetry properties of the amplitudes implied by Hermiticity and the time-reversal transformation. 
In particular, the relations in Eq.~\eqref{e:HTR_cons} imply that the amplitudes $\widetilde{A}_3$, $\widetilde{A}_4$, and $\widetilde{A}_8$ are odd in $\xi$.  This symmetry behavior, plus the requirement of a well-defined forward limit of the matrix element in Eq.~\eqref{helicity_para}, leads us to conclude that these three amplitudes vanish for $\xi = 0$. 
In the analysis of the lattice data presented in this work, we first kept those amplitudes as nonzero and indeed found them numerically to be compatible with zero.  
More discussion can be found in Sec.~\ref{sec:results}.

We will now establish connections between the light-cone GPDs and the amplitudes. As certain quantities depend on the kinematic frame, it becomes crucial to differentiate and specify the relevant frame. To achieve this distinction, we employ superscripts $s$ and $a$ to denote the symmetric and asymmetric frames, respectively. We note that in the symmetric frame, the momentum transfer is equally distributed between the initial and final states, while any other distribution is considered asymmetric.
After substituting $\mu =+$ in Eq.~(\ref{helicity_para}), we can apply a basis transformation to relate the $\widetilde{A}_i$'s in the resulting expression to the GPDs in Eq.~(\ref{e:GPD_def_pos}):
\begin{align}
\widetilde{H}(z\cdot P^{s/a}, z \cdot \Delta^{s/a}, (\Delta^{s/a})^2)  & = \widetilde{A}_2 + (P^{s/a, +} z^-) \widetilde{A}_6 + (\Delta^{s/a,+} z^-) \widetilde{A}_8 \nonumber \\[0.2cm]
& = \widetilde{A}_2 + (P^{s/a} \cdot z) \widetilde{A}_6 + (\Delta^{s/a} \cdot z) \widetilde{A}_8 \, , \label{e:H_LC}\\[0.2cm]
\widetilde{E} (z\cdot P^{s/a}, z \cdot \Delta^{s/a}, (\Delta^{s/a})^2) & =  2 \, \dfrac{P^{s/a,+} }{\Delta^{s/a,+}} \widetilde{A}_3 + 2 \widetilde{A}_5 \nonumber \\[0.2cm]
& =  2 \, \dfrac{P^{s/a} \cdot z}{\Delta^{s/a} \cdot z} \widetilde{A}_3 + 2 \widetilde{A}_5 \, ,
\label{e:E_LC}
\end{align}
where the $\widetilde{A}_i$'s are evaluated at $z^2 = 0$.
We emphasize that in the aforementioned equations, we have expressed the kinematic variables multiplying the amplitudes using Lorentz-invariant scalars. 
It is crucial to note that this particular re-writing is unique. 
Furthermore, as evident now, these equations exhibit the property of Lorentz invariance, which guarantees their validity and applicability across different reference frames.
It is worth noting that due to the prefactor $\Delta^+$, $\widetilde{E}$ drops out of the parametrization, Eq.~(\ref{e:GPD_def_pos}), at $\xi = 0$. Upon first look the expression for $\widetilde{E}$, Eq.~(\ref{e:E_LC}), seems only valid for $\xi \neq 0$.
However, based on symmetry arguments (as discussed before and given in Appendix~\ref{s:symmetries}), $\widetilde{A}_3$ is odd in $\xi$ and vanishes at $\xi = 0$. 
One can therefore reliably determine the zero-skewness limit of $\widetilde{A}_3/\xi$ (and of $\widetilde{E}$) by calculating $\widetilde{A}_3$ for nonzero $\xi$ and extrapolating the r.h.s.~of Eq.~\eqref{e:E_LC} to $\xi = 0$.
Note that a recent work has unveiled the possibility of obtaining a glimpse into $\widetilde{E}$ at $\xi =0$ by studying a specific twist-3 GPD~\cite{Bhattacharya:2023nmv}.

Now, we shift our focus to quasi-GPDs. 
As highlighted in Ref.~\cite{Bhattacharya:2022aob}, one plausible approach to define the quasi-GPDs is by starting from the Lorentz-invariant light-cone definitions in Eqs.~(\ref{e:H_LC}) and $(\ref{e:E_LC})$ to incorporate $z^2 \neq 0$. Throughout our discussions, we will refer to this definition as the Lorentz-invariant (LI) quasi-GPD:
\begin{align}
\widetilde{\mathcal{H}}(z\cdot P^{s/a}, z \cdot \Delta^{s/a}, (\Delta^{s/a})^2, z^2) & = \widetilde{A}_2 + (P^{s/a} \cdot z) \widetilde{A}_6 + (\Delta^{s/a} \cdot z) \widetilde{A}_8 \, , \label{e:H_LI}\\[0.2cm]
\widetilde{\mathcal{E}} (z\cdot P^{s/a}, z \cdot \Delta^{s/a}, (\Delta^{s/a})^2, z^2) & =  2 \, \dfrac{P^{s/a} \cdot z}{\Delta^{s/a} \cdot z} \widetilde{A}_3 + 2 \widetilde{A}_5 \, ,
\label{e:E_LI}
\end{align}
where now the $\widetilde{A}_i$ are evaluated at $z^2 \neq 0$~\footnote{It is important to emphasize that the choice of basis does not impact the final results for LI quasi-GPDs, as discussed in Appendix~\ref{s:another_basis}, where we explore an alternative basis for the axial-vector case and demonstrate the basis-independence of the LI quasi-GPD result.}. 
In simple words, this definition of the quasi-GPD is based on the same functional form in terms of the $\widetilde{A}_i$ as the light-cone GPD (see Eq.~(\ref{e:H_LC}) and Eq.~(\ref{e:E_LC})). 
One finds that this definition is given by an operator which combines $(\gamma^{0}, \gamma^1, \gamma^2) \gamma_5$, rather than the conventional operator $\gamma^3\gamma_5$ (further explained in the subsequent paragraph and sections). Because of this operator structure, a different matching coefficient is required compared to the one used for $\gamma^3 \gamma_5$. Specifically, one can disregard the contributions from the operators $\gamma^1 \gamma_5$ and $\gamma^2 \gamma_5$ as they are relatively suppressed by a factor of $1/(P^3)^2$. In our numerical results, we chose to implement the matching for $\gamma^3\gamma_5$ (see Ref.~\cite{Liu:2019urm} and also Eq.~(\ref{eq:kernel})) due to the unavailability of the matching kernel for $\gamma^0 \gamma_5$ in the literature. Consequently, this necessitates a new calculation for both the matching and the renormalization of the $\gamma^0 \gamma_5$ operator, along with addressing its mixing effects.
Note also that the difference between results with matching for $\gamma^0 \gamma_5$ and $\gamma^3 \gamma_5$ is finite and is expected numerically to be very small.

We now turn to the set of quasi-GPDs already introduced in Eq.~\eqref{e:qGPD_def_pos}. By setting $\mu=3$ in Eq.~(\ref{helicity_para}), we can perform a change of basis to transform the resulting expression and establish a mapping between the $\widetilde{A}_i$ and the quasi-GPDs defined in Eq.~(\ref{e:qGPD_def_pos}). The relations are as follows:
\begin{align}
\label{eq:qH3}
\widetilde{\mathcal{H}}_3 (z, P^{s/a}, \Delta^{s/a})  & = \widetilde{A}_2 - z^3 P^{3,s/a} \widetilde{A}_6 - m^2 (z^3)^2 \widetilde{A}_7 - z^3 \Delta^{3,s/a} \widetilde{A}_8 \, ,\\[0.2cm]
\label{eq:qE3}
\widetilde{\mathcal{E}}_3 (z, P^{s/a}, \Delta^{s/a})  & =  2 \, \dfrac{P^{3,s/a}}{\Delta^{3,s/a}} \widetilde{A}_3 + 2 m^2 \dfrac{z^3}{\Delta^{3,s/a}} \widetilde{A}_4 + 2 \widetilde{A}_5 \, .
\end{align}
Similar to the light-cone case, it is important to note that there are no divergences arising from the terms $\widetilde{A}_3/\xi$ and $\widetilde{A}_4/\xi$ as $\xi \rightarrow 0$ in Eq.~(\ref{eq:qE3}). 
We iterate that the reason for the well-behaved $\xi \rightarrow 0$ limit is the fact that these amplitudes are odd in $\xi$ and vanish for $\xi \to 0$ (see also Appendix~\ref{s:symmetries}). 
Moreover, if one intends to calculate the value of $\widetilde{\mathcal{E}}$ at $\xi = 0$ using Eq.~(\ref{eq:qE3}), extrapolation from nonzero $\xi$ values becomes necessary.

Now, let us examine the frame (in)dependence of Eqs.~(\ref{eq:qH3})-(\ref{eq:qE3}). It is noteworthy that these equations hold true in both symmetric and asymmetric frames, which is why we have refrained from using explicit ``s/a" superscripts to denote the GPDs.
The kinematical prefactor of the amplitudes can (again) be uniquely expressed through Lorentz scalars:
\begin{align}
\widetilde{\mathcal{H}}_3 (z\cdot P^{s/a}, z \cdot \Delta^{s/a}, (\Delta^{s/a})^2, z^2)  & = \widetilde{A}_2 + (P^{s/a} \cdot z) \widetilde{A}_6 + m^2 z^2 \widetilde{A}_7 + (\Delta^{s/a} \cdot z) \widetilde{A}_8 \, , \label{e:Hq_LI}\\[0.2cm]
\widetilde{\mathcal{E}}_3 (z\cdot P^{s/a}, z \cdot \Delta^{s/a}, (\Delta^{s/a})^2, z^2)  & =  2 \, \dfrac{P^{s/a} \cdot z}{\Delta^{s/a} \cdot z} \widetilde{A}_3 + 2 m^2 \dfrac{z^2}{\Delta^{s/a} \cdot z} \widetilde{A}_4 + 2 \widetilde{A}_5 \label{e:Eq_LI} \, .
\end{align}
In our previous study~\cite{Bhattacharya:2022aob}, we emphasized the frame-dependence of the conventional definitions of unpolarized quasi-GPDs that employ $\gamma^0$. 
In contrast, we observe that the helicity quasi-GPDs defined through $\gamma^3 \gamma_5$ are the same in the symmetric and non-symmetric frames.  
This can be understood since the two frames are connected through a transverse boost which preserves the 3-component. 
By applying the same reasoning, it can be inferred that quasi-GPDs defined using $\gamma^0\gamma_5$ will exhibit frame dependence. 
Furthermore, given the ability to reformulate the traditional definition involving $\gamma^3 \gamma_5$ in a Lorentz-invariant manner, it is clear that it emerges as an additional contender for a Lorentz-invariant definition. 
Consequently, this example explicitly demonstrates the lack of uniqueness in Lorentz-invariant definitions for quasi-GPDs. 
The distinction between the Lorentz-invariant definitions presented in Eqs.~(\ref{e:H_LI})-(\ref{e:E_LI}) and Eqs.~(\ref{e:Hq_LI})-(\ref{e:Eq_LI}) can be attributed to terms proportional to $z^2$ associated with the amplitudes $\widetilde{A}_4$ and $\widetilde{A}_7$.

We repeat that the two sets of quasi-GPDs discussed above are not equivalent, as they differ in the contributing amplitudes and both explicit and implicit power corrections.
Henceforth, when referring to power corrections, we specifically denote corrections that are proportional to $z^2$. 
The additional amplitudes in Eqs.~(\ref{e:Hq_LI})-(\ref{e:Eq_LI}) can be interpreted as contamination arising from explicit power corrections, which could potentially be suppressed by considering higher values of the momentum. 
One may therefore speculate that Eqs.~(\ref{e:H_LI})-(\ref{e:E_LI}) converge faster compared to Eqs.~(\ref{e:Hq_LI})-(\ref{e:Eq_LI}). 
However, it is essential to acknowledge that the amplitudes themselves also contain implicit power corrections, so the above statement should be examined case by case. 
(The presence of additional amplitudes could potentially mitigate the implicit power corrections inherent in the amplitudes stated in Eqs.~(\ref{e:H_LI})-(\ref{e:E_LI}).) 
Ultimately, the actual convergence of the different quasi-GPD definitions is determined by the underlying non-perturbative dynamics. 
Therefore, it is important to perform numerical comparisons to assess the convergence behavior of these definitions and gain insights into the relative magnitude of power corrections in each case.

We conclude this section by briefly discussing the symmetry properties of $\widetilde{H}$ (and $\widetilde{E}$) in position space, as these properties play a crucial role in leveraging symmetries to improve statistical precision in lattice calculations. 
The Hermiticity constraint provides the symmetries of GPDs under the transformation $P^3 \rightarrow -P^3$ for a fixed value of $z^3$. 
Similarly, for a fixed $P^3$ and with $\Delta \rightarrow -\Delta$, the Hermiticity constraint unveils the symmetries of GPDs under the transformation $z^3 \rightarrow -z^3$. 
Notably, we find that the real part of $\widetilde{H}$ satisfies $\widetilde{H}(-P^3) = +\widetilde{H}(P^3)$ (and likewise for $\widetilde{E}$). 
Furthermore, we observe $\widetilde{H}(-z^3) = +\widetilde{H}(z^3)$ (and likewise for $\widetilde{E}$). 
(The imaginary parts of the GPDs satisfy the same constraints as their real parts, with the exception of a negative sign.) 
Finally, we would like to reiterate that a comprehensive analysis of the symmetries at the level of amplitudes is provided in Appendix~\ref{s:symmetries}.

\section{Lattice Calculation}
\subsection{Methodology}
\label{sec:mat_meth}

In this section, we present a synopsis of the methodology for the lattice QCD calculation of proton matrix elements using the axial-vector operator. 
All expressions here are presented in Euclidean space, where we use lower indices in $P$ and $\Delta$ to avoid confusion in the expressions given previously in Minkowski space. 
The goal of this calculation is twofold: (a) compare the Lorentz invariant amplitudes extracted from different frames; (b) present results for the $\widetilde{H}$ GPD at multiple values of $t$. We note that our calculation is performed at zero skewness and, thus, $\widetilde{E}$ is inaccessible from the matrix elements.
The two frames we employ are the symmetric
\bea
\label{eq:pf_symm}
\vec{p}^{\,s}_f=\vec{P} + \frac{\vec{\Delta}}{2} = \left(+\frac{\Delta_1}{2},+\frac{\Delta_2}{2},P_3\right)\,, \quad
\vec{p}^{\,s}_i=\vec{P} - \frac{\vec{\Delta}}{2}= \left(-\frac{\Delta_1}{2},-\frac{\Delta_2}{2},P_3\right)\,,
\eea
and an asymmetric in which the final state does not contain the momentum transfer
\bea
\label{eq:pf_nonsymm}
\vec{p}^{\,a}_f=\vec{P} =  \left(0,0,P_3\right) \,,\quad
\vec{p}^{\,a}_i=\vec{P} - \vec{\Delta} =  \left(-\Delta_1,-\Delta_2,P_3\right)\,.
\eea
In the above equations, a factor of $\frac{2\pi}{L}$ ($L$: spatial extent of the lattice) is implied in $\vec{\Delta}$ and $P_3$. Note that $\vec{\Delta}$ is the same in both frames, however, $-t\equiv\vec{\Delta}^2 - (E_f - E_i)^2$ differs due to the term containing the energies.
In each kinematic frame, we parametrize the lattice matrix elements using the trace
\begin{equation}
 K\, {\rm Tr}\Big[\Gamma_\kappa \, \left(\frac{-ip_f \hspace*{-0.33cm}\slash \,\,+m}{2m} \right) \, F^{[\gamma_\mu \gamma_5]} \left(\frac{-ip_i \hspace*{-0.27cm}\slash \,\,+m}{2m} \right) \Big]\,, \quad \mu, \kappa=0,1,2,3\,,
 \label{eq:tr} 
\end{equation}
where $F^{[\gamma_\mu \gamma_5]}$ is given in Eq.~\eqref{helicity_para}. 
We use four parity projectors; the unpolarized, $\Gamma_0$, and the three polarized, $\Gamma_k$, defined as
\begin{eqnarray}
\Gamma_0 &=& \frac{1}{4} \left(1 + \gamma_0\right)\,, \\
\Gamma_k &=& \frac{1}{4} \left(1 + \gamma_0\right) i \gamma_5 \gamma_k\,, \quad k=1,2,3\,.
\end{eqnarray}
$K$ is a kinematic factor that has been obtained based on the normalization of the proton state, 
\begin{equation}
K = \frac{2 m^2}{\sqrt{E_f E_i (E_f + m) (E_i + m)}}\,.
\end{equation}
As we will demonstrate below, the combination of the four projectors and the four directions of the axial-vector operator can disentangle all eight $\widetilde{A}_i$ for any kinematic setup. 

Next, we focus on the ground-state contribution to the matrix elements that we will denote as $\Pi^{s/a}_\mu(\Gamma_\kappa)$ ($\mu, \kappa:\,0,\,1,\,2,\,3$).
We note that the operator $\gamma_j \gamma_5$ ($j\neq 3$) has a finite mixing under renormalization for lattice regularizations with chiral symmetry breaking~\cite{Constantinou:2017sej,Alexandrou:2017ypw,Chen:2017mie,Green:2017xeu}. 
Such mixing is not included in the renormalization analysis in this calculation, as it would require the matrix elements of the tensor operator. 
However, the effect is found to be small for the twisted mass formulation with a clover term~\cite{Constantinou:2017sej,Alexandrou:2017ypw}.
The general expressions for Eq.~\eqref{eq:tr} in the symmetric frame for zero skewness are
\begin{eqnarray}
\label{eq:Pi0G0_s}
  \Pi^s_0(\Gamma_0) &=& 0  \hspace*{0.75cm}\\[3ex]
\label{eq:Pi0G1_s}
  \Pi^s_0(\Gamma_1) &=& K \,     \left(
\frac{E  \Delta_1  ( E +m)}{4 m^3}\, \widetilde{A}_3 
    \right)\\[3ex]
\label{eq:Pi0G2_s}
  \Pi^s_0(\Gamma_2) &=& K\,  \left(
\frac{E  \Delta_2  ( E +m)}{4 m^3}\, \widetilde{A}_3   
   \right)\\[3ex]
  \Pi^s_0(\Gamma_3) &=&  K\,  \left(
-\frac{ 
     P_3  ( E +m)}{2 m^2}\,\widetilde{A}_2 -\frac{  E^2 z
    ( E +m)}{2 m^2}\,\widetilde{A}_6     \right) 
    \label{eq:Pi1G0_s}
\end{eqnarray}
\begin{eqnarray}    
  \Pi^s_1(\Gamma_0) &=& K\,     \left(
-\frac{2    E   \Delta_2  z
    \left( E  ( E +m)- P_3^2\right)}{4 m^3}\,\widetilde{A}_1 -\frac{ 
     P_3   \Delta_2 }{4 m^2}  \,\widetilde{A}_2 
    \right)\\[3ex]
\label{eq:Pi1G1_s}
  \Pi^s_1(\Gamma_1) &=& i\,K\,  \left(
-\frac{ E 
     P_3   \Delta_2^2 z}{4m^3}\,\widetilde{A}_1   +\frac{  \left(4 m
    ( E +m)+ \Delta_2^2\right)}{8 m^2}\,\widetilde{A}_2 -\frac{   \Delta_1^2
    ( E +m)}{4 m^3}\,\widetilde{A}_5 
    \right)\\[3ex]
\label{eq:Pi1G2_s}
  \Pi^s_1(\Gamma_2) &=& i\,K\,  \left(
\frac{ 
     E   P_3   \Delta_1   \Delta_2  z}{4m^3}\,\widetilde{A}_1 -\frac{  \Delta_1 
     \Delta_2 }{8 m^2}\,\widetilde{A}_2  -\frac{  \Delta_1   \Delta_2  ( E +m)}{4
    m^3} \,\widetilde{A}_5  
    \right)\\[3ex]
\label{eq:Pi1G3_s}
    \Pi^s_1(\Gamma_3) &=& i\,K\,   \left(
    \frac{  E  \Delta_1  z ( E +m)}{2 m^2}\,\widetilde{A}_8  
  \right) \\[3ex]
  \label{eq:Pi2G0_s}
  \Pi^s_2(\Gamma_0) &=& K\,  \left(
\frac{2  
     E   \Delta_1  z \left( E 
    ( E +m)- P_3^2\right)}{4 m^3}\,\widetilde{A}_1 +\frac{ P_3 
     \Delta_1 }{4 m^2}\, \widetilde{A}_2  
  \right)\\[3ex]
  \label{eq:Pi2G1_s}
    \Pi^s_2(\Gamma_1) &=& i\,K\, \left(
\frac{   E   P_3 
     \Delta_1   \Delta_2  z}{4m^3}\,\widetilde{A}_1 -\frac{  \Delta_1   \Delta_2 }{8
    m^2}\, \widetilde{A}_2 -\frac{  \Delta_1   \Delta_2  ( E +m)}{4
    m^3}\, \widetilde{A}_5 
    \right)\\[3ex]
\label{eq:Pi2G2_s}
  \Pi^s_2(\Gamma_2) &=& i\,K\,  \left(
-\frac{  E 
     P_3   \Delta_1^2 z}{4m^3}\,\widetilde{A}_1  +\frac{ \left(4 m
    ( E +m)+ \Delta_1^2\right)}{8 m^2}\, \widetilde{A}_2 -\frac{ \Delta_2^2
    ( E +m)}{4 m^3}\, \widetilde{A}_5  
  \right)\\[3ex]
\label{eq:Pi2G3_s}
    \Pi^s_2(\Gamma_3) &=& i\,K\, \left(
    \frac{  E  \Delta_2  z ( E +m)}{2 m^2}\,\widetilde{A}_8     \right) \\[3ex]
\label{eq:Pi3G0_s}
  \Pi^s_3(\Gamma_0) &=& 0\\[3ex]
\label{eq:Pi3G1_s}
  \Pi^s_3(\Gamma_1) &=& i\,K\, \left(
   -\frac{  P_3   \Delta_1  ( E +m)}{4
    m^3}\, \widetilde{A}_3 +
\frac{   \Delta_1  z ( E +m)}{4
    m}\,\widetilde{A}_4
  \right)\\[3ex]
\label{eq:Pi3G2_s}
  \Pi^s_3(\Gamma_2) &=& i\,K\,  \left(
  -\frac{ P_3   \Delta_2  ( E +m)}{4
    m^3}\, \widetilde{A}_3  +
\frac{   \Delta_2  z
    ( E +m)}{4 m}\, \widetilde{A}_4
  \right)\\[3ex]
\label{eq:Pi3G3_s}
  \Pi^s_3(\Gamma_3) &=& iK\,  \left(
\frac{  E  ( E +m)}{2
    m^2}\, \widetilde{A}_2 +\frac{  E   P_3  z ( E +m)}{2
    m^2}\,\widetilde{A}_6  -\frac{1}{2}  E  z^2 ( E +m)\,  \widetilde{A}_7 
  \right)\,,
\end{eqnarray}

\noindent
where $K$ simplifies to $2 m^2/(E(E+m))$, due to $E_i=E_f\equiv E$ in the symmetric frame when $\xi=0$.
It is interesting to observe that $\widetilde{A}_7$ appears only in Eq.~\eqref{eq:Pi3G3_s}. 
In general, we find fourteen nonzero equations, some of which are linearly dependent. 
For instance, Eqs.~\eqref{eq:Pi0G1_s} and \eqref{eq:Pi0G2_s} have the same numerical value besides a multiplicative factor of $\Delta_1$ and $\Delta_2$, respectively. 
Still, there are eight linearly independent matrix elements that allow one to disentangle all amplitudes $\widetilde{A}_i$. 
Another observation is that the amplitudes $\widetilde{A}_3$, $\widetilde{A}_4$, $\widetilde{A}_8$ are decoupled from the other ones.
This is an important aspect, as these amplitudes are expected to be zero at $\xi=0$ due to theoretical arguments.
In Sec.~\ref{sec:results}, we will comment more about how one can incorporate this information into the analysis. 

\newpage
The trace algebra of Eq.~\eqref{eq:tr} in the asymmetric frame of Eq.~\eqref{eq:pf_nonsymm} leads to more complicated kinematic coefficients mainly because $E_i \ne E_f$, as well as the lack of symmetry between $p_f$ and $p_i$. At zero skewness, we obtain
\begin{eqnarray}
\label{eq:Pi0G0_ns}
  \Pi^a_0(\Gamma_0) &=& 0  \\[3ex]
  \Pi^a_0(\Gamma_1) &=& K\,     \Bigg(
\frac{ ( E_f +m)  \Delta_1 }{4 m^2}\, \widetilde{A}_2 +\frac{ 
    ( E_f + E_i ) ( E_f +m)  \Delta_1 }{8 m^3}\, \widetilde{A}_3
    +\frac{ 
    ( E_f - E_i ) ( E_f +m)  \Delta_1 }{4 m^3}\, \widetilde{A}_5
    +\frac{ 
    ( E_f + E_i )  P_3  z  \Delta_1 }{8 m^2}\, \widetilde{A}_6\nonumber \\[1ex]
   && \qquad
    +\frac{
    ( E_f - E_i )  P_3  z  \Delta_1 }{4 m^2}\, \widetilde{A}_8 
    \Bigg)\\[3ex]
\label{eq:Pi0G2_ns}
  \Pi^a_0(\Gamma_2) &=& K\,  \Bigg(
\frac{  ( E_f +m)  \Delta_2 }{4 m^2}\,\widetilde{A}_2 +\frac{ 
    ( E_f + E_i ) ( E_f +m)  \Delta_2 }{8 m^3}\, \widetilde{A}_3 +\frac{ 
    ( E_f - E_i ) ( E_f +m)  \Delta_2 }{4 m^3}\, \widetilde{A}_5
    +\frac{ 
    ( E_f + E_i )  P_3  z  \Delta_2 }{8
    m^2}\, \widetilde{A}_6\nonumber \\[1ex]
   && \qquad  +\frac{ 
    ( E_f - E_i )  P_3  z  \Delta_2 }{4 m^2}\, \widetilde{A}_8
   \Bigg)\\[3ex]
  \Pi^a_0(\Gamma_3) &=& K\,  \Bigg(
 -\frac{ 
    ( E_f + E_i +2 m)  P_3 }{4 m^2}\, \widetilde{A}_2+\frac{ 
    \left( E_i^2- E_f^2\right)  P_3 }{8 m^3}\, \widetilde{A}_3 -\frac{  P_3 
    ( E_f - E_i )^2}{4 m^3}\, \widetilde{A}_5
    -\frac{ 
    ( E_f + E_i )^2 ( E_f +m) z}{8 m^2}\, \widetilde{A}_6\nonumber \\[1ex]
   && \qquad -\frac{  ( E_f + E_i )
    ( E_f +m) z ( E_f - E_i )}{4 m^2}\, \widetilde{A}_8
   \Bigg)\\[3ex]
  \Pi^a_1(\Gamma_0) &=& K\,     \Bigg(
\frac{    E_f 
    ( E_f - E_i -2 m) ( E_f +m)  \Delta_2  z}{4m^3}\,\widetilde{A}_1-\frac{ 
     P_3   \Delta_2 }{4 m^2}\, \widetilde{A}_2
    \Bigg)\\[3ex]
  \Pi^a_1(\Gamma_1) &=& i\,K\,  \Bigg(
-\frac{  E_f   P_3   \Delta_2^2
    z}{4m^3}\, \widetilde{A}_1 +\frac{ 
    \left(( E_f +m) ( E_i +m)- P_3^2\right)}{4
    m^2}\,\widetilde{A}_2  +\frac{  
    ( E_f +m)  \Delta_1^2}{8 m^3}\,\widetilde{A}_3
    -\frac{ ( E_f +m)
     \Delta_1^2}{4 m^3}\, \widetilde{A}_5\nonumber \\[1ex]
   && \qquad  +\frac{   P_3  z  \Delta_1^2}{8
    m^2}\, \widetilde{A}_6 -\frac{  P_3  z  \Delta_1^2}{4 m^2}\, \widetilde{A}_8 
    \Bigg)  \\[3ex]
  \Pi^a_1(\Gamma_2) &=& i\,K\,  \Bigg(
  \frac{   E_f   P_3   \Delta_1  z
     \Delta_2 }{4m^3}\, \widetilde{A}_1 +
\frac{ ( E_f +m)  \Delta_1   \Delta_2 }{8
    m^3}\, \widetilde{A}_3 -\frac{   ( E_f +m)  \Delta_1   \Delta_2 }{4
    m^3}\,\widetilde{A}_5+\frac{   P_3   \Delta_1  z  \Delta_2 }{8
    m^2}\, \widetilde{A}_6
     -\frac{   P_3   \Delta_1  z  \Delta_2 }{4
    m^2}\, \widetilde{A}_8
    \Bigg) \qquad\\[3ex]
    \Pi^a_1(\Gamma_3) &=& i\,K\,  \Bigg(
-\frac{    P_3   \Delta_1 }{4 m^2}\,\widetilde{A}_2+\frac{  
    ( E_i - E_f )  P_3   \Delta_1 }{8 m^3}\,\widetilde{A}_3 +\frac{ 
    ( E_f - E_i )  P_3   \Delta_1 }{4 m^3}\, \widetilde{A}_5
    -\frac{  
    ( E_f + E_i ) ( E_f +m) z  \Delta_1 }{8 m^2}\,\widetilde{A}_6\nonumber \\[1ex]
   && \qquad+\frac{ 
    ( E_f + E_i ) ( E_f +m) z  \Delta_1 }{4
    m^2} \, \widetilde{A}_8
  \Bigg) \\[3ex]
  \Pi^a_2(\Gamma_0) &=& K\,  \Bigg(
-\frac{ E_f 
    ( E_f - E_i -2 m) ( E_f +m)  \Delta_1  z}{4m^3}\, \widetilde{A}_1 + \frac{ 
     P_3   \Delta_1 }{4 m^2}\, \widetilde{A}_2  
  \Bigg)\\[3ex]
  \Pi^a_2(\Gamma_1) &=& i\,K\,  \Bigg(
\frac{  E_f   P_3   \Delta_1  z
     \Delta_2 }{4m^3}\, \widetilde{A}_1 + \frac{ ( E_f +m)
     \Delta_1   \Delta_2 }{8 m^3}\, \widetilde{A}_3 -\frac{( E_f +m)  \Delta_1 
     \Delta_2 }{4 m^3}\, \widetilde{A}_5  +\frac{   P_3   \Delta_1  z  \Delta_2 }{8
    m^2}\, \widetilde{A}_6
   -\frac{  P_3   \Delta_1  z  \Delta_2 }{4
    m^2}\, \widetilde{A}_8 
  \Bigg)
\end{eqnarray}
\begin{eqnarray}
  \Pi^a_2(\Gamma_2) &=& i\,K\,  \Bigg(
-\frac{ 
     E_f   P_3  z  \Delta_1^2}{4m^3}\, \widetilde{A}_1+\frac{ ( E_f +m)
     \Delta_2^2}{8 m^3}\, \widetilde{A}_3 +\frac{ \left(( E_f +m)
    ( E_i +m)- P_3^2\right)}{4 m^2}\, \widetilde{A}_2 -\frac{ ( E_f +m)  \Delta_2^2}{4
    m^3}\, \widetilde{A}_5  \nonumber \\[1ex]
   && \qquad+\frac{ P_3 
     \Delta_2^2 z}{8 m^2}\, \widetilde{A}_6  -\frac{  P_3   \Delta_2^2 z}{4
    m^2}\, \widetilde{A}_8 
  \Bigg)\\[3ex]
\label{eq:Pi2G3_ns}
    \Pi^a_2(\Gamma_3) &=& i\,K\,  \Bigg(
-\frac{   P_3   \Delta_2 }{4
    m^2}\,\widetilde{A}_2+\frac{  ( E_i - E_f )  P_3   \Delta_2 }{8
    m^3}\, \widetilde{A}_3 +\frac{  ( E_f - E_i )  P_3   \Delta_2 }{4
        m^3}\, \widetilde{A}_5
        -\frac{  ( E_f + E_i ) ( E_f +m) z  \Delta_2 }{8
    m^2}\, \widetilde{A}_6\nonumber \\[1ex]
   && \qquad+\frac{  ( E_f + E_i ) ( E_f +m) z  \Delta_2 }{4
    m^2} \, \widetilde{A}_8
     \Bigg) \\[3ex]
  \Pi^a_3(\Gamma_0) &=& 0 \\[3ex]
  \Pi^a_3(\Gamma_1) &=& i\,K\, \Bigg(
 -\frac{  \Delta_1   P_3 }{4
    m^2}\, \widetilde{A}_2 -\frac{ ( E_f +m)  \Delta_1 
     P_3 }{4 m^3}\,\widetilde{A}_3  +\frac{ ( E_f +m)  \Delta_1  z}{4 m}\, \widetilde{A}_4 -\frac{   \Delta_1  z  P_3^2}{4 m^2}\, \widetilde{A}_6+\frac{1}{4}
          \Delta_1  z^2  P_3\,\widetilde{A}_7   
    \Bigg)\\[3ex]
  \Pi^a_3(\Gamma_2) &=& i\,K\,  \Bigg(
-\frac{  \Delta_2 
     P_3 }{4 m^2}\, \widetilde{A}_2 -\frac{ 
    ( E_f +m)  \Delta_2   P_3 }{4 m^3}\, \widetilde{A}_3+\frac{ ( E_f +m)  \Delta_2  z}{4
    m}\, \widetilde{A}_4  -\frac{  \Delta_2  z  P_3^2}{4
    m^2}\, \widetilde{A}_6 +\frac{1}{4}   \Delta_2  z^2  P_3\, \widetilde{A}_7  
  \Bigg) \qquad \\[3ex]
\label{eq:Pi3G3_ns}
  \Pi^a_3(\Gamma_3) &=&  i\,K\,  \Bigg(
\frac{ 
    ( E_f + E_i ) ( E_f +m)}{4 m^2}\, \widetilde{A}_2 + \frac{ ( E_f - E_i )  P_3^2}{4
    m^3}\, \widetilde{A}_3  +\frac{ ( E_i - E_f ) z  P_3 }{4 m}\, \widetilde{A}_4 
    +\frac{ ( E_f + E_i ) ( E_f +m) z  P_3 }{4
    m^2}\, \widetilde{A}_6 
     \nonumber \\[1ex]
   && \qquad -\frac{1}{4}
   ( E_f + E_i ) ( E_f +m) z^2\,  \widetilde{A}_7  
  \Bigg)\,.
\end{eqnarray}
To summarize our findings, Eqs.~\eqref{eq:Pi0G0_s} - \eqref{eq:Pi3G3_s} and Eqs.~\eqref{eq:Pi0G0_ns} - \eqref{eq:Pi3G3_ns} are sufficient to disentangle the $\widetilde{A}_i$ in the symmetric and asymmetric frame, respectively. This task can be done analytically by inverting the equations, which, however, leads to very complicated general expressions; it is practically more convenient to implement a numerical inversion of the $8\times8$ system for each value of $P$ and $\vec{\Delta}$. Here, we give the expressions for $\widetilde{A}_i$ using $\vec{\Delta}= (\Delta,0,0)$ as an example.
We use a superscript $s$ and $a$ in the matrix elements to differentiate between the two frames; $\widetilde{A}_i$ are frame-independent and do not carry such an index.
The expressions for the symmetric frame take the form
 \begin{eqnarray}
 \label{eq:A1_s}
\widetilde{A}_1&=& \frac{\,i\, m P_3 }{2 z
    (E+m) \left(E^2-P_3^2\right)}\Pi_2^s(\Gamma_2) +\frac{m 
    \left(E^2+E m-P_3^2\right)}{ z \Delta 
    (E+m) \left(E^2-P_3^2\right)} \Pi_2^s(\Gamma_0)\,,\\[3ex]
\widetilde{A}_2&=&
    \frac{E P_3 \Delta}{2 (E+m)
    \left(E^2-P_3^2\right)} \Pi_2^s(\Gamma_0)+\frac{i\,E 
     \left(P_3^2-E
    (E+m)\right)}{(E+m) (E-P_3)
    (E+P_3)}\Pi_2^s(\Gamma_2)\,,\\[3ex]
\widetilde{A}_3&=& \frac{2 m    }{\Delta}\Pi_0^s(\Gamma_1)\,,\\[3ex]
\widetilde{A}_4&=& \frac{2 P_3
    }{z \Delta }\Pi_0^s(\Gamma_1)-\frac{2 \,i\, E 
    }{z \Delta m}\Pi_3^s(\Gamma_1)\,,\\[3ex]
\widetilde{A}_5&=& -\frac{2\,i\, E  m^2
     \left(E^2+E m-P_3^2\right)}{\Delta^2
    (E+m) \left(E^2-P_3^2\right)}\Pi_2^s(\Gamma_2)+\frac{E m^2 P_3}{\Delta (E+m)
    \left(E^2-P_3^2\right)} \Pi_2^s(\Gamma_0)+\frac{2  \,i\,E m
    }{\Delta^2}\Pi_1^s(\Gamma_1)\,,
\end{eqnarray}
\begin{eqnarray}
\widetilde{A}_6&=& \frac{i\, P_3
    \left(E^2+E m-P_3^2\right)}{E z
    (E+m) \left(E^2-P_3^2\right)} \Pi_2^s(\Gamma_2)+\frac{P_3^2
    \Delta }{2 E z (E+m)
    \left(P_3^2-E^2\right)}\Pi_2^s(\Gamma_0)-\frac{1}{E
    z}\Pi_0^s(\Gamma_3)\,,\\[3ex]
\widetilde{A}_7&=&\frac{i\,  \left(P_3^2-E
    (E+m)\right)}{E m^2 z^2 (E+m)}\Pi_2^s(\Gamma_2)+\frac{P_3
    \Delta }{2 E m^2 z^2 (E+m)}\Pi_2^s(\Gamma_0)-\frac{P_3
    }{E m^2 z^2}\Pi_0^s(\Gamma_3)+\frac{i\, }{m^2
    z^2}\Pi_3^s(\Gamma_3)\,,\\[3ex]
\widetilde{A}_8&=& -\frac{i\, }{z \Delta }\Pi_1^s(\Gamma_3)\,,
\end{eqnarray}
 and for the asymmetric frame at $\vec{\Delta}=(\Delta,0,0)$, one obtains
 \begin{eqnarray}
z\,\widetilde{A}_1&=& \frac{2 (E_f-E_i-2 m) m^3 }{
    E_f (E_i+m) \left(E_f^2{-}E_i E_f{-}2 m^2\right)
      \Delta}\frac{\Pi^a_2(\Gamma_0)}{K}-\frac{2 i\, m^3 P_3}{ E_f
    (E_f+m) (E_i+m) \left(E_f^2{-}E_i E_f{-}2
    m^2\right) }\frac{\Pi^a_2(\Gamma_2)}{K} \,,\\[3ex]
\widetilde{A}_2&=& \frac{2 P_3 \Delta  
    m^2}{(E_f+m) (E_i+m) \left(2 m^2+E_f
    (E_i-E_f)\right)}\frac{\Pi^a_2(\Gamma_0)}{K}+\frac{2 \,i\, (E_f-E_i-2 m)
     m^2}{(E_i+m) \left(2 m^2+E_f
    (E_i-E_f)\right)}\frac{\Pi^a_2(\Gamma_2)}{K} \,,\\[3ex]
\widetilde{A}_3&=&  \frac{2 (E_f+E_i)
     m^3}{E_f^2 (E_i+m) \Delta }\frac{\Pi^a_0(\Gamma_1)}{K}+\frac{2 P_3
     m^3}{E_f^2 (E_f+m) (E_i+m)}\frac{\Pi^a_0(\Gamma_3)}{K}+\frac{2
    \,i\,  m^3}{E_f^2 (E_i+m)}\frac{\Pi^a_1(\Gamma_1)}{K}\nonumber \\[1.2ex]
&&+\frac{2\,i\,
    (E_i-E_f)  P_3 m^3}{E_f^2
    (E_f+m) (E_i+m) \Delta }\frac{\Pi^a_1(\Gamma_3)}{K}  \,,\\[3ex]
z\,\widetilde{A}_4&=& \frac{2
    (E_f+E_i) P_3  m^2}{E_f^2 (E_i+m)
      \Delta}\frac{\Pi^a_0(\Gamma_1)}{K}+\frac{2 (E_f-m)  m^2}{E_f^2
    (E_i+m)}\frac{\Pi^a_0(\Gamma_3)}{K} -\frac{2 \,i\, (E_f-E_i) (E_f-m)
     m^2}{E_f^2 (E_i+m)   \Delta}\frac{\Pi^a_1(\Gamma_3)}{K}\nonumber \\[1.2ex]
&&+\frac{2 \,i\, P_3  m^2}{E_f^2
    (E_i+m) }\frac{\Pi^a_1(\Gamma_1)}{K} -\frac{2 \,i\, 
    (E_f+E_i) m^2}{E_f (E_i+m)
      \Delta}\frac{\Pi^a_3(\Gamma_1)}{K}-\frac{2 \,i\, P_3 m^2}{E_f
    (E_f+m) (E_i+m) }\frac{\Pi^a_3(\Gamma_3)}{K}  \,,\\[3ex]
\widetilde{A}_5&=&  -\frac{2 (E_f+E_i)
    P_3  m^4}{E_f (E_f+m) (E_i+m)
    \left(E_f^2{-}E_i E_f{-}2 m^2\right) \Delta }\frac{\Pi^a_2(\Gamma_0)}{K}+\frac{(E_f+E_i) 
    m^3}{E_f^2 (E_i+m) \Delta }\frac{\Pi^a_0(\Gamma_1)}{K} \nonumber \\[1.2ex]
&&+\frac{2
    \,i\, (E_f-E_i-2 m)  m^4}{E_f
    (E_f-E_i) (E_i+m) \left(E_f^2-E_i
    E_f-2 m^2\right)}\frac{\Pi^a_2(\Gamma_2)}{K}+\frac{P_3 
    m^3}{E_f^2 (E_f+m) (E_i+m)}\frac{\Pi^a_0(\Gamma_3)}{K}\nonumber \\[1.2ex]
&&-\frac{ i\, (E_f+E_i)
    m^3}{E_f^2 (E_f-E_i)
    (E_i+m)}\frac{\Pi^a_1(\Gamma_1)}{K}+\frac{i\,(E_f+E_i) P_3 
    m^3}{E_f^2 (E_f+m) (E_i+m) \Delta }\frac{\Pi^a_1(\Gamma_3)}{K} \,,\\[3ex]
z\,\widetilde{A}_6&=& \frac{2 (E_i-E_f) P_3  m^2}{E_f^2
    (E_f+m) (E_i+m)   \Delta}\frac{\Pi^a_0(\Gamma_1)}{K}-\frac{2 
    m^2}{E_f^2 (E_i+m) }\frac{\Pi^a_0(\Gamma_3)}{K} +\frac{2\,i\, (E_f-E_i)  
    m^2}{E_f^2 (E_i+m)   \Delta}\frac{\Pi^a_1(\Gamma_3)}{K}\nonumber \\[1.2ex]
&&+\frac{2  \,i\,(E_i-E_f)
    P_3 m^2}{E_f^2 (E_f+E_i) (E_f+m)
    (E_i+m) }\frac{\Pi^a_1(\Gamma_1)}{K} -\frac{4 (E_f-E_i)
    (E_f-m)  m^2}{E_f (E_i+m)
        \left(E_f^2{-}E_i E_f{-}2 m^2\right)   \Delta}\frac{\Pi^a_2(\Gamma_0)}{K}\nonumber \\[1.2ex]
&&+\frac{4
    \,i\, (E_f-E_i-2 m) P_3 m^2}{E_f
    (E_f+E_i) (E_i+m) \left(E_f^2-E_i
    E_f-2 m^2\right) } \frac{\Pi^a_2(\Gamma_2)}{K} \,,
\end{eqnarray}
\begin{eqnarray}
z^2\,\widetilde{A}_7&=&  -\frac{2 (E_f-E_i)
    (E_f-m) }{E_f^2 (E_i+m) \Delta 
    }\frac{\Pi^a_0(\Gamma_1)}{K}-\frac{2 P_3 }{E_f^2 (E_i+m)}\frac{\Pi^a_0(\Gamma_3)}{K}-\frac{2\,i\, 
    (E_f-E_i) (E_f-m) }{E_f^2
    (E_f+E_i) (E_i+m)}\frac{\Pi^a_1(\Gamma_1)}{K}\nonumber \\[1.2ex]
&&+\frac{2  \,i\,(E_f-E_i)
    P_3 }{E_f^2 (E_i+m) \Delta 
    }\frac{\Pi^a_1(\Gamma_3)}{K}+\frac{2 (E_i-E_f) P_3 }{E_f
    (E_f+m) (E_i+m) \Delta}\frac{\Pi^a_2(\Gamma_0)}{K}+\frac{2 \,i\,
    (E_f-E_i-2 m) }{E_f (E_f+E_i)
    (E_i+m)}\frac{\Pi^a_2(\Gamma_2)}{K}\nonumber \\[1.2ex]
&&+\frac{2 \,i\, (E_f-E_i) P_3
    }{E_f (E_f+m) (E_i+m)  \Delta}\frac{\Pi^a_3(\Gamma_1)}{K}+\frac{2
    \,i\, }{E_f (E_i+m) } \frac{\Pi^a_3(\Gamma_3)}{K}\,,\\[3ex]
z\,\widetilde{A}_8&=&    \frac{(E_i-E_f) P_3 m^2}{E_f^2
    (E_f+m) (E_i+m)   \Delta}\frac{\Pi^a_0(\Gamma_1)}{K} -\frac{
    m^2}{E_f^2 (E_i+m) }\frac{\Pi^a_0(\Gamma_3)}{K}+\frac{i\, P_3 
    m^2}{E_f^2 (E_f+m) (E_i+m)
    }\frac{\Pi^a_1(\Gamma_1)}{K}\nonumber \\[1.2ex]
&&-\frac{i\, (E_f+E_i) m^2}{E_f^2
    (E_i+m)   \Delta}\frac{\Pi^a_1(\Gamma_3)}{K} 
\end{eqnarray}

The use of the amplitudes $\widetilde{A}_i$ is a pathway to extracting the quasi-GPDs using lattice data from any kinematic frame. 
Here, we present two approaches to relate the $\widetilde{A}_i$ to the quasi-GPDs: (a) the standard $\gamma_3\gamma_5$ operator (Eq.~\eqref{eq:FH_s}); (b) an alternative Lorentz-invariant definition (Eq.~\eqref{eq:FH_LI}). 
Our focus is on zero skewness, which only gives access to the $\widetilde{H}$ GPD; the kinematic coefficient of $\widetilde{E}$ in Eq.~\eqref{e:qGPD_def_pos} becomes zero due to the factor $\Delta_3$. 
For the same reason, $\widetilde{E}$ does not appear in the parametrization of the matrix elements in the forward limit. 
In fact, to obtain $\widetilde{E}$-GPD at $t=0$, one must parametrize its $t$ dependence, and, similarly, its estimate at zero skewness could be obtained by a fit using $\xi\neq0$ values.
Below, we give the relation between the quasi-GPD of $\widetilde{H}$ at zero skewness using the $\gamma_3 \gamma_5$ definition. 
We note that the standard definition of Eq.~\eqref{eq:FH_s} is Lorentz invariant, and therefore, it is the same in both frames, as discussed in Sec.~\ref{s:decomposition}. 
At zero skewness, one obtains 
\begin{eqnarray}
\label{eq:FH_s}
\widetilde{\cal H}_3(\widetilde{A}_i^{s/a};z) &=& \widetilde{A}_2 + z P_3 \widetilde{A}_6 - m^2 z^2 \widetilde{A}_7 \,.
\end{eqnarray}
For simplicity, we only show two arguments for $\widetilde{\cal H}_3$, that is, $\widetilde{A}_i$ to indicate the frame used in the calculation and $z$ to explicitly show that the relation for quasi-GPDs is given in coordinate space. 
To keep the expressions compact, we suppress the arguments of the amplitudes $\widetilde{A}_i$. 
It is useful to rewrite Eq.~\eqref{eq:FH_s} in terms of matrix elements in the symmetric frame for the special case $\vec{\Delta}=(\Delta,0,0)$, for which we find
\begin{eqnarray}
\label{eq:FHs_Pi}
\widetilde{\cal H}_3(\widetilde{A}_i^s;z)  &=& -i \,\Pi^s_3(\Gamma_3)\,.
\end{eqnarray}
As expected, Eq.~\eqref{eq:FHs_Pi} is the usual expression extracted from the matrix elements of the $\gamma_3\gamma_5$ operator previously used for the helicity GPDs~\cite{Alexandrou:2020zbe}. 
With the $\widetilde{A}_i$ being frame-invariant, one can use either $A^s_i$ or $A^a_i$ in Eq.~\eqref{eq:FH_s}; calculating $A^a_i$ is computationally less costly and, thus, more optimal for lattice QCD calculations.

An alternative approach to extract the light-cone GPDs is through a Lorentz-invariant definition of choice for the quasi-GPDs $\widetilde{\cal H}$ and $\widetilde{\cal E}$, as given in Eqs.~\eqref{e:Hq_LI} - \eqref{e:Eq_LI}, where $\widetilde{\cal H}(\widetilde{A}_i^s;z) =\widetilde{\cal H}(\widetilde{A}_i^a;z)$ and $\widetilde{\cal E}(\widetilde{A}_i^s;z) =\widetilde{\cal E}(\widetilde{A}_i^a;z)$, by construction. 
The expression for $\widetilde{\cal H}$ at zero skewness simplifies giving
\begin{eqnarray}
\label{eq:FH_LI}
\widetilde{\cal H}(\widetilde{A}_i^{s/a};z)   &= & \widetilde{A}_2 +  z P_3 \widetilde{A}_6  \,.
\end{eqnarray}
For completeness, we provide the expressions of $\widetilde{\cal H}$ using matrix elements in each frame. 
As above, we use as an example the case $\vec{\Delta}=(\Delta,0,0)$ to write $\widetilde{\cal H}$ in terms of matrix elements, that is
\begin{eqnarray}
\label{eq:Hq_LI_s}
\widetilde{\cal H}(\widetilde{A}_i^s;z)  &=& \frac{P_3  \Delta
    }{2 E^ 2+2 E  m}\Pi^s_2(\Gamma_0)+ i\,
    \frac{P_3 ^2 - E  (E +m)}{E  (E +m)}\Pi^s_2(\Gamma_2) -\frac{P_3 }{E } \Pi^s_0(\Gamma_3)\,, 
\end{eqnarray}
    \begin{eqnarray}
\label{eq:Hq_LI_a}
\widetilde{\cal H}(\widetilde{A}_i^a;z)  &=& \frac{2\,  i\, m^2 P_3 ^2 (E_i -E_f )
}{E_f^ 2 (E_f +E_i ) (E_f +m)
    (E_i +m)}\frac{\Pi^a_1(\Gamma_1)}{K}-\frac{2 \, i\, m^2 (E_f -E_i -2
    m) \left(E_f  (E_f +E_i )-2 P_3 ^2\right)}{E_f 
    (E_f +E_i ) (E_i +m) \left(E_f^ 2-E_f 
    E_i -2 m^2\right)}\frac{\Pi^a_2(\Gamma_2)}{K} \nonumber \\[1.2ex]
    && +\frac{2 \, i\, m^2 P_3 
    (E_f -E_i )}{E_f^ 2 \Delta
    (E_i +m)}\frac{\Pi^a_1(\Gamma_3)}{K}+\frac{2 m^2 P_3 ^2 (E_i -E_f )}{E_f^ 2 \Delta (E_f +m) (E_i +m)}\frac{\Pi^a_0(\Gamma_1)}{K} \nonumber \\[1.2ex]
    && +\frac{2
    m^2 P_3) \left(2 (E_f -E_i )
    \left(m^2-E_f^ 2\right)-E_f  \Delta^2\right)}{E_f 
    \Delta (E_f +m) (E_i +m) \left(E_f^ 2-E_f 
    E_i -2 m^2\right)}\frac{\Pi^a_2(\Gamma_0)}{K}-\frac{2 m^2 P_3}{E_f^ 2
    (E_i +m)} \frac{\Pi^a_0(\Gamma_3)}{K}\,,
\end{eqnarray}
This alternative definition of $\widetilde{\cal H}$ can be interpreted as the construction of a new operator that is a combination of $\gamma_\mu\gamma_5$ with $\mu=0,\,1,\,2,\,3$, as given in the example of Eqs.~\eqref{eq:Hq_LI_s} - \eqref{eq:Hq_LI_a}.
We note, however, that in the case of the helicity, the matrix elements $\gamma_{k}\gamma_5$ with $k=0,1,2$ ($k\ne3$) have finite mixing in lattice regularization~\cite{Constantinou:2017sej}, which affects $\widetilde{\cal H}$.
In Sec.~\ref{sec:results}, we will compare the two definitions of $\widetilde{\cal H}_3$ and $\widetilde{\cal H}$ and discuss their merits.

\subsection{Computational setup}

The proton matrix elements entering Eq.~\eqref{eq:tr} ($F^{[\gamma_\mu\gamma_5]}$) use a non-local axial-vector operator containing spatially-separated quark fields in the $\hat{z}$ direction. 
The Wilson line and the momentum boost are also along the $\hat{z}$ direction.
The matrix elements have momentum transfer between the initial and final state,  $\vec{\Delta}=\vec{p_f}-\vec{p_i}$, and can be written as 
\begin{equation}
\label{eq:ME}
F^{[\gamma_\mu\gamma_5]}(\Gamma_\kappa,z,p_f,p_i)\equiv \langle N(p_f)|\bar\psi\left(z\right) \gamma_\mu \gamma_5 {\cal W}(0,z)\psi\left(0\right)|N(p_i)\rangle\,, \quad \mu, \kappa: 0,1,2,3\,.
\end{equation}
$|N(p_i)\rangle$ and $|N(p_f)\rangle$ are the initial (source) and final (sink) states of the proton, while the remaining variables are defined previously.
We use momentum smearing~\cite{Bali:2016lva} to improve the overlap with the proton ground state and suppress gauge noise; Ref.~\cite{Alexandrou:2016jqi} demonstrated that the method is essential for non-local operators.
It was also found that the statistical noise is $z$-dependent and reduces by a factor of 4-5 in the real part and 2-3 in the imaginary part of the quasi-GPDs calculated in a previous work~\cite{Alexandrou:2020zbe}. 
In addition, we use five steps of stout smearing~\cite{Morningstar:2003gk} to the gauge links of the operator with parameter $\rho=0.15$, to further suppress gauge noise, as demonstrated in Refs.~\cite{Alexandrou:2016ekb,Alexandrou:2020sml}. We note that the stout smearing changes both the matrix elements and the renormalization function, but the renormalized matrix elements should remain independent of the stout smearing. Indeed, in Ref.~\cite{Alexandrou:2019lfo} it was examines the effect of the number of stout smearing steps (0, 5, 10, 15, 20) and shows that the renormalized matrix elements are stout-smearing independent; the test was performed at $\vec{\Delta}=0$ and a physical pion mass ensemble. The same conclusions were reached in the case of the gluon PDF~\cite{Delmar:2023agv} calculated using the same ensemble as this work.
The matrix element is extracted from the ratio
\begin{equation}
\label{eq:ratio}
R_\mu (\Gamma_\kappa, z, p_f, p_i; t_s, \tau) = \frac{C^{\rm 3pt}_\mu (\Gamma_\kappa, z, p_f, p_i; t_s, \tau)}{C^{\rm 2pt}(\Gamma_0, p_f;t_s)} \sqrt{\frac{C^{\rm 2pt}(\Gamma_0, p_i, t_s-\tau)C^{\rm 2pt}(\Gamma_0, p_f, \tau)C^{\rm 2pt}(\Gamma_0, p_f, t_s)}{C^{\rm 2pt}(\Gamma_0, p_f, t_s-\tau)C^{\rm 2pt}(\Gamma_0, p_i, \tau)C^{\rm 2pt}(\Gamma_0, p_i, t_s)}}\,,
 \end{equation}
where $C^{\rm 2pt}$ and $C^{\rm 3pt}$, are the two- and three-point correlation functions. 
$\tau$ is the current insertion time, and $t_s$ is the source-sink time separation; the source is taken at zero timeslice.
As shown in Table~\ref{tab:stat}, we implement all kinematically equivalent momenta that lead to the same value of $p_i^2$, $p_f^2$. 
Thus, to increase statistical accuracy, we average $C^{\rm 2pt}$ for all possible values. 
We extract the ground-state contribution to $F^{[\gamma_\mu\gamma_5]}$ from $R_\mu$ by taking a plateau fit with respect to $\tau$ in a region of convergence. Here, we indicate the ground state by $\Pi_\mu(\Gamma_\kappa)$, and their decomposition is given in Eqs.~\eqref{eq:Pi0G0_s} - \eqref{eq:Pi3G3_ns}. 
For simplicity, the dependence on $z$, $p_f$, and $p_i$ is not shown explicitly in the matrix elements $\Pi^j(\Gamma_\kappa)$.

\medskip
The calculation is performed on a gauge ensemble of $N_f=2+1+1$ twisted-mass fermions, including a clover term~\cite{Alexandrou:2018egz}. 
The gluon part of the action is Iwasaki-improved.
The volume of the ensemble is $32^3 \times 64$, and its lattice spacing, $a$, is 0.093 fm.
The quark masses correspond to a pion mass of 260 MeV. The ensemble parameters are given in Table~\ref{tab:params}.
 \begin{table}[h!]
\centering
\renewcommand{\arraystretch}{1.2}
\renewcommand{\tabcolsep}{6pt}
  \begin{tabular}{| l| c | c | c | c | c  | c | c |}
  \hline
    \multicolumn{8}{|c|}{Parameters} \\
    \hline
 Ensemble   & $\beta$ & $a$ [fm] & volume $L^3\times T$ & $N_f$ & $m_\pi$ [MeV] &
$L m_\pi$ & $L$ [fm]\\
    \hline
cA211.32 & 1.726 & 0.093  & $32^3\times 64$  & $u, d, s, c$ & 260
& 4 & 3.0 \\
    \hline
    \end{tabular}
  \caption{\small Parameters of the ensemble used in this work.}
  \label{tab:params}
  \vspace*{0.2cm}
  \end{table}
Using this ensemble, we obtain the matrix elements at a source-sink time separation of $t_s=10 a = 0.934$ fm, a choice made to effectively manage statistical uncertainties in the matrix elements. 
The study of excited states via calculations of multiple time separations lies outside the scope of the current project. 
Details regarding the statistics of the calculation in both the symmetric and asymmetric frames are provided in Table~\ref{tab:stat}. 
In summary, for $P_3=1.25$ GeV we analyze three different values of $-t$ in the symmetric frame. 
These are complemented by seven values of $-t$ in the asymmetric frame, which significantly enhances computational efficiency. 
The majority of the values are within the range $-t \in [0.17 - 1.50]$ GeV. To examine dependence on the momentum boost, we focus on $-t^s=0.69$ GeV$^2$, where we use three values of $P_3$, that is 0.83, 1.25, and 1.67 GeV. 
Notably, the asymmetric frame offers a computationally advantageous approach, as it enables the acquisition of multiple values of $\vec{\Delta}$ within the same computational cost. To elaborate, while each value of $t$ in the symmetric frame requires a separate calculation, the data production in the asymmetric frame is divided into two groups: one for $(\pm\Delta_x,0,0)$ and its permutations and another for $(\pm\Delta_x,\pm\Delta_y,0)$ and their permutations.

\begin{table}[h!]
\begin{center}
\renewcommand{\arraystretch}{1.9}
\begin{tabular}{lcccc|cccc}
\hline
frame & $P_3$ [GeV] & $\quad \mathbf{\Delta}$ $[\frac{2\pi}{L}]\quad$ & $-t$ [GeV$^2$] & $\quad \xi \quad $ & $N_{\rm ME}$ & $N_{\rm confs}$ & $N_{\rm src}$ & $N_{\rm tot}$\\
\hline
N/A       & $\pm$1.25 &(0,0,0)  &0   &0   &2   &329  &16  &10528 \\
\hline
symm      & $\pm$0.83 &($\pm$2,0,0), (0,$\pm$2,0)  &0.69   &0   &8   &67 &8  &4288 \\
symm      & $\pm$1.25 &($\pm$2,0,0), (0,$\pm$2,0)  &0.69   &0   &8   &249 &8  &15936 \\
symm      & $\pm$1.67 &($\pm$2,0,0), (0,$\pm$2,0)  &0.69   &0   &8   &294 &32  &75264 \\
symm      & $\pm$1.25 &$(\pm 2,\pm 2,0)$           &1.38   &0   &16   &224 &8  &28672 \\
symm      & $\pm$1.25 &($\pm$4,0,0), (0,$\pm$4,0)  &2.77   &0   &8   &329 &32  &84224 \\
\hline
asymm  & $\pm$1.25 &($\pm$1,0,0), (0,$\pm$1,0)  &0.17   &0   &8   &269 &8  &17216\\
asymm      & $\pm$1.25 &$(\pm 1,\pm 1,0)$       &0.34   &0   &16   &195 &8  &24960 \\
asymm  & $\pm$1.25 &($\pm$2,0,0), (0,$\pm$2,0)  &0.65   &0   &8   &269 &8  &17216\\
asymm      & $\pm$1.25 &($\pm$1,$\pm$2,0), ($\pm$2,$\pm$1,0) &0.81   &0   &16   &195 &8  &24960 \\
asymm  & $\pm$1.25 &($\pm$2,$\pm$2,0)          &1.24    &0   &16  &195 &8   &24960\\
asymm  & $\pm$1.25 &($\pm$3,0,0), (0,$\pm$3,0)  &1.38   &0   &8   &269 &8  &17216\\
asymm      & $\pm$1.25 &($\pm$1,$\pm$3,0), ($\pm$3,$\pm$1,0)  &1.52   &0   &16   &195 &8  &24960 \\
asymm  & $\pm$1.25 &($\pm$4,0,0), (0,$\pm$4,0)  &2.29   &0   &8   &269 &8  &17216\\
\hline
\end{tabular}
\caption{\small Statistics for the symmetric and asymmetric frame matrix elements are shown. The momentum unit $2\pi/L$ is 0.417 GeV. $N_{\rm ME}$, $N_{\rm confs}$, $N_{\rm src}$ and $N_{\rm total}$ are the number of matrix elements, configurations, source positions per configuration and total statistics, respectively.}
\label{tab:stat}
\end{center}
\end{table}

\medskip
The benefits of acquiring the data summarized in Table~\ref{tab:stat} are three-fold, as it allows: 
\begin{itemize}
\item[--] Comparison of results on the three values of $P_3$ at fixed $-t$ to assess $P_3$ dependence; 
\item[--] Comparison of the estimates for $\widetilde{A}_i$ in the two frames using $-t^s=0.69$ GeV$^2$ and $-t^a=0.65$ GeV$^2$. 
\item[--] Extraction of the $-t$ dependence of the GPDs and apply parametrizations. 
\end{itemize}
In comparing $\widetilde{A}_i$ between frames, it's worth noting that although $t^s$ and $t^a$ are not precisely identical, they are closely aligned, differing by only 5\%, which allows for a meaningful and reliable comparison.
\newpage
\section{Lattice results}
\label{sec:results}

\subsection{Comparison of kinematic frame}

In this section, we focus on the setup with $P_3=\pm1.25$ GeV and $\vec{\Delta}=\{\frac{2\pi}{L}(\pm2,0,0), \frac{2\pi}{L}(0,\pm2,0)\}$, implemented in both the symmetric and asymmetric frames. This setup gives $-t^s=0.69$ GeV$^2$ in the symmetric frame, and $-t^a=0.65$ GeV$^2$ in the asymmetric frame. 
Our first goal is to compare the $\widetilde{A}_i$ between the two frames, in a similar fashion as our previous work for the unpolarized GPDs~\cite{Bhattacharya:2022aob}. 
Once agreement is established between the $\widetilde{A}_i$ from the two frames, all data from Table~\ref{tab:stat} will be analyzed to extract the $-t$ dependence of GPDs. 

\begin{figure}[h!]
    \centering
    \includegraphics[scale=0.39]{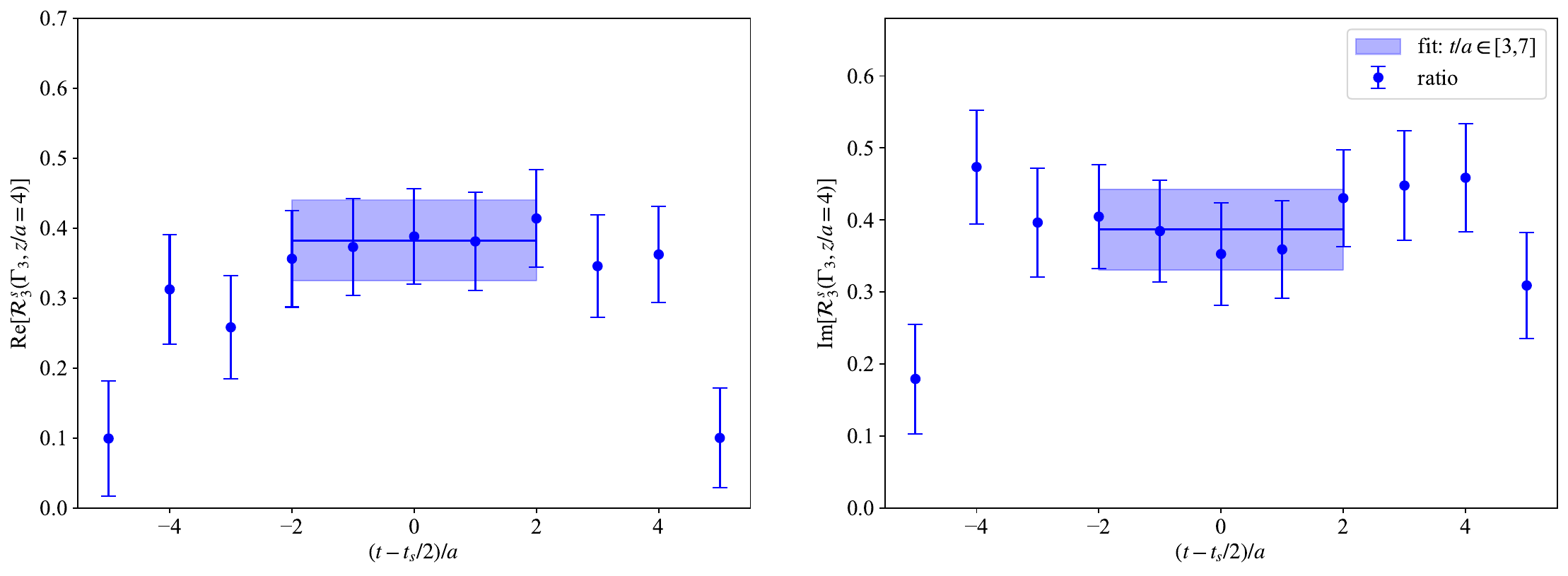}
    \vspace*{-0.4cm}
\caption{\small{The ratio of Eq.~\eqref{eq:ratio} for $\Pi_3(\Gamma_3)$ in the symmetric frame for $|P_3|=1.25$ GeV and $\vec{\Delta}=\frac{2\pi}{L}(2,0,0)$. The left (right) panel corresponds to the real (imaginary) part.}}
    \label{fig:R3G3}
\end{figure}

\begin{figure}[h!]
    \centering
    \includegraphics[scale=0.39]{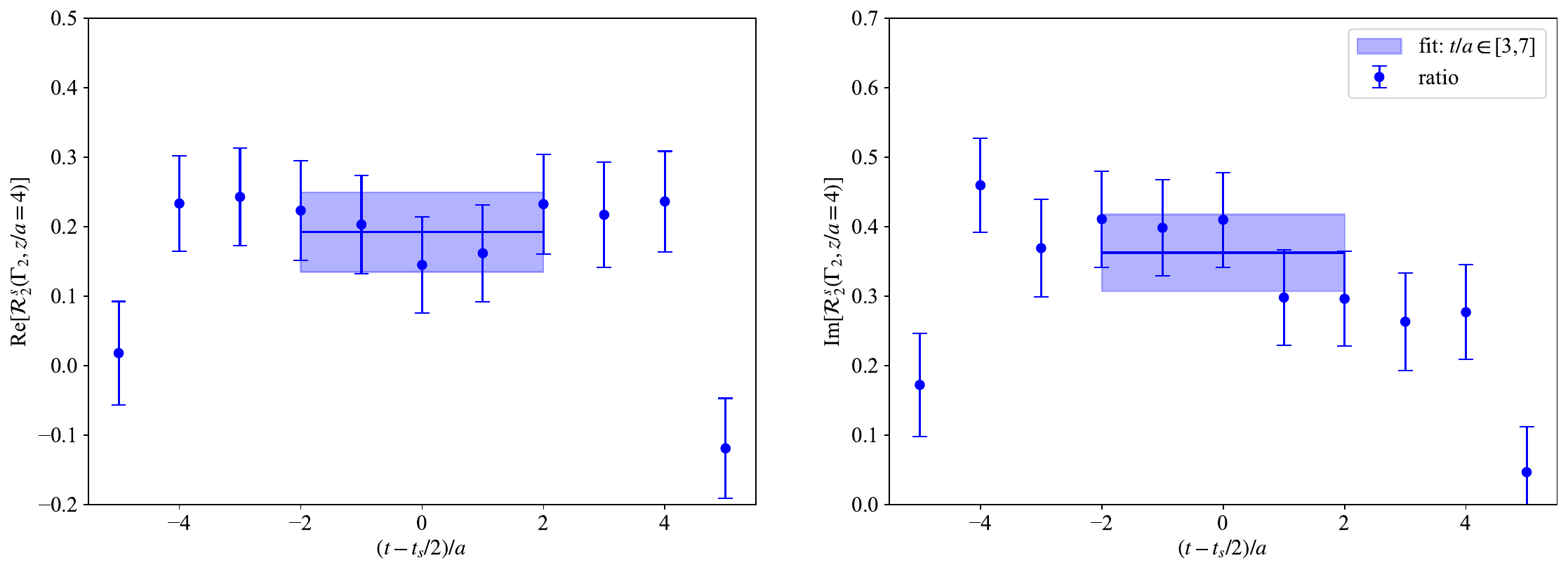}
    \vspace*{-0.4cm}
\caption{\small{The ratio of Eq.~\eqref{eq:ratio} for $\Pi_2(\Gamma_2)$ in the symmetric frame for $|P_3|=1.25$ GeV and $\vec{\Delta}=\frac{2\pi}{L}(2,0,0)$. The left (right) panel corresponds to the real (imaginary) part.}}
    \label{fig:R2G2}
\end{figure}

Before presenting the matrix elements, we show the ratio of Eq.~\eqref{eq:ratio} for two representative cases, $R_3(\Gamma_3)$ and $R_2(\Gamma_2)$ and $P_3=1.25$ GeV and $\vec{\Delta}=\frac{2\pi}{L}(2,0,0)$. For better clarity, we use the symmetric frame in which the data have less statistical fluctuations. We choose $[3a-7a]$ for the fit with respect to the insertion time
We remind the reader that the data of Figs.~\ref{fig:R3G3} - \ref{fig:R2G2} are only one of the eight kinematically equivalent cases. These are averaged at the amplitude level according to the symmetry properties of the latter.

\begin{figure}[h!]
    \centering
    \includegraphics[scale=0.38]{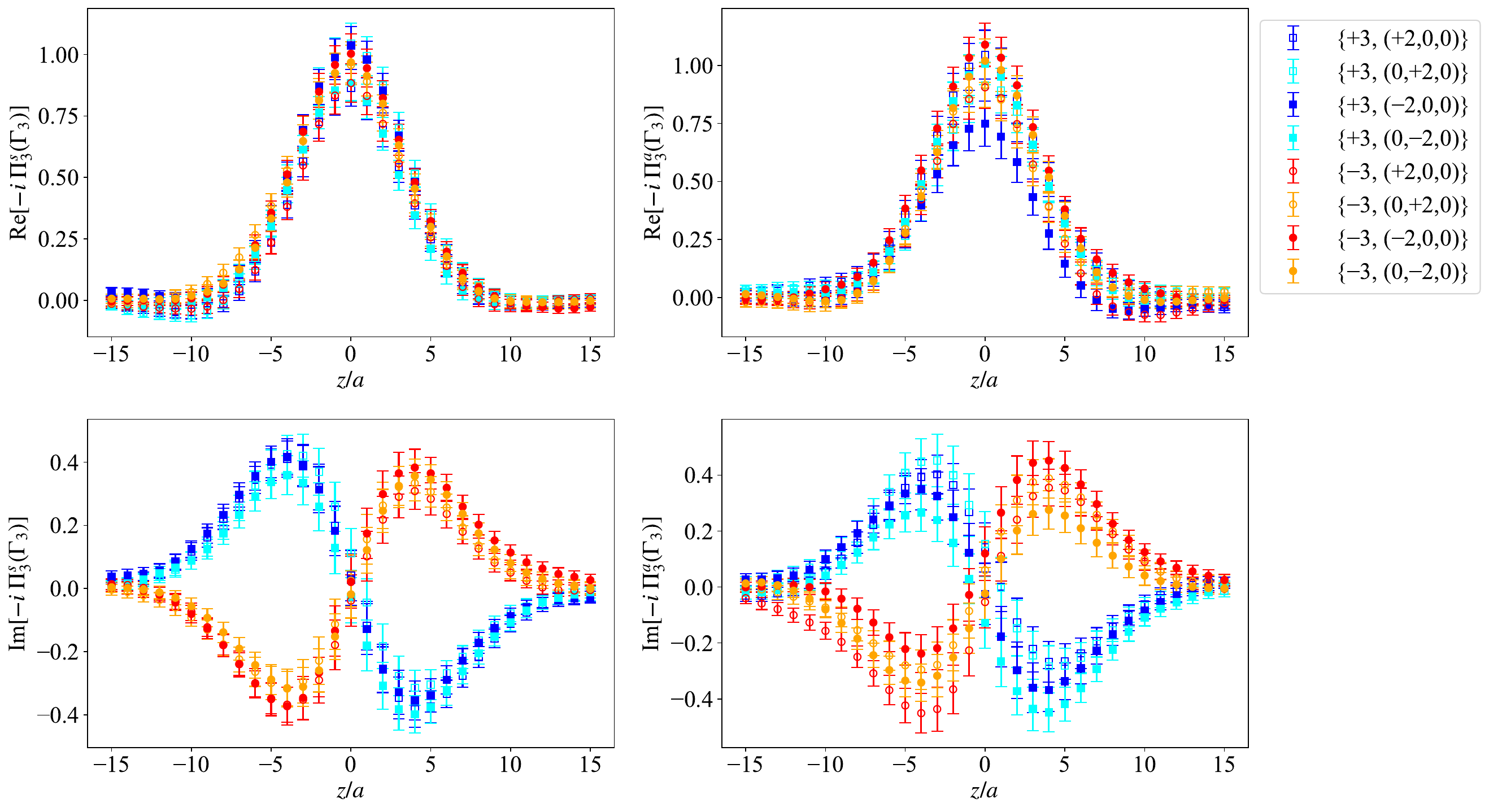}
    \vspace*{-0.4cm}
\caption{\small{Bare matrix elements $\Pi_3(\Gamma_3)$ in the symmetric frame (left) and in the asymmetric frame (right), for $|P_3|=1.25$ GeV and $-t=0.69$ GeV$^2$ ($-t=0.65$ GeV$^2$) for the symmetric (asymmetric) frame. The top (bottom) panel corresponds to the real (imaginary) part. The notation in the legend is $\{P_3,\vec{\Delta}\}$ in units of $2\pi/L$.}}
    \label{fig:Pi3G3}
\end{figure}

\begin{figure}[h!]
    \centering
    \includegraphics[scale=0.32]{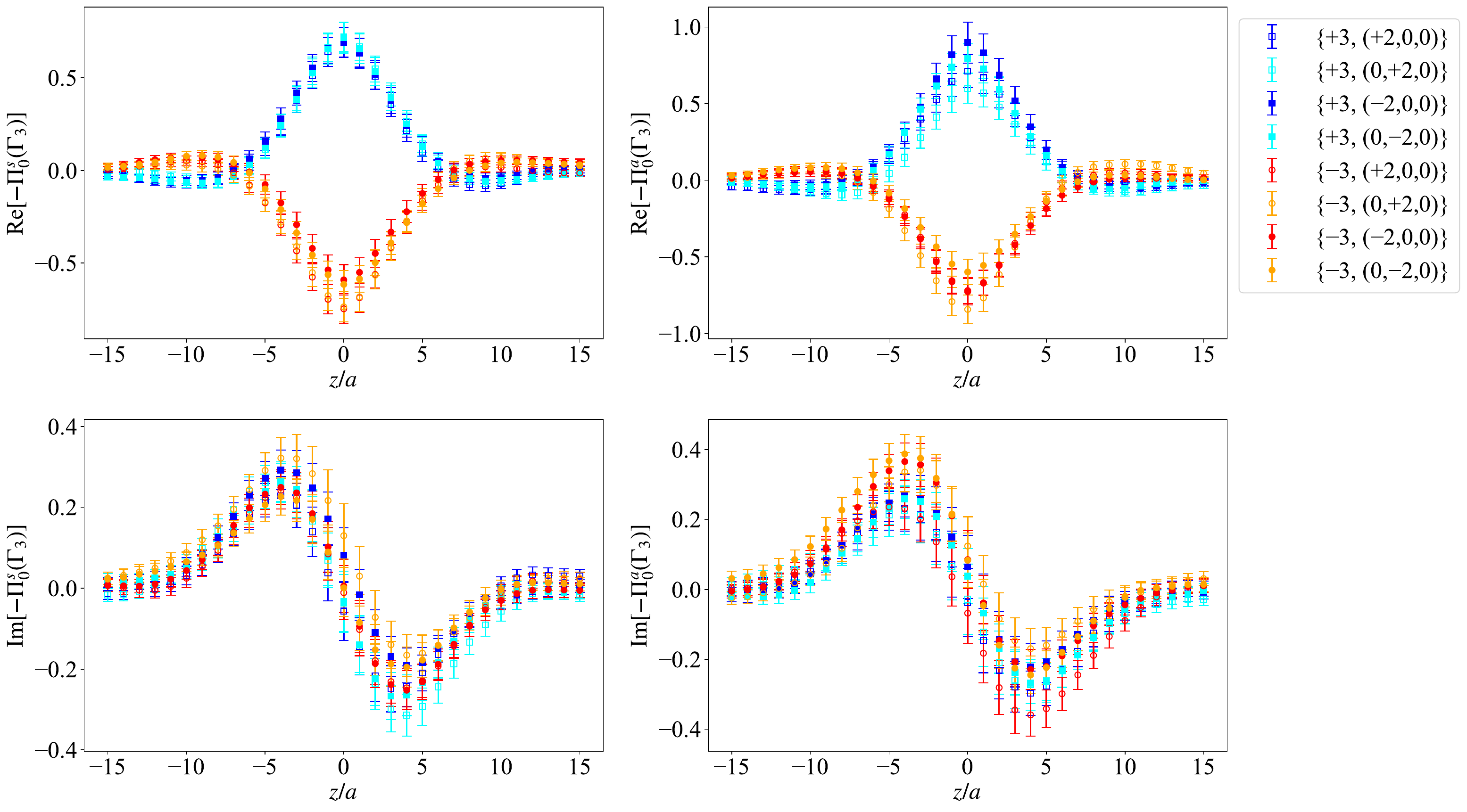}
    \vspace*{-0.35cm}
    \caption{\small Bare matrix elements $\Pi_0(\Gamma_3)$ in the symmetric frame (left) and in the asymmetric frame (right). The notation is the same as Fig.~\ref{fig:Pi3G3}.}
    \label{fig:Pi0G3}
\end{figure}

\begin{figure}[h!]
    \centering
    \includegraphics[scale=0.38]{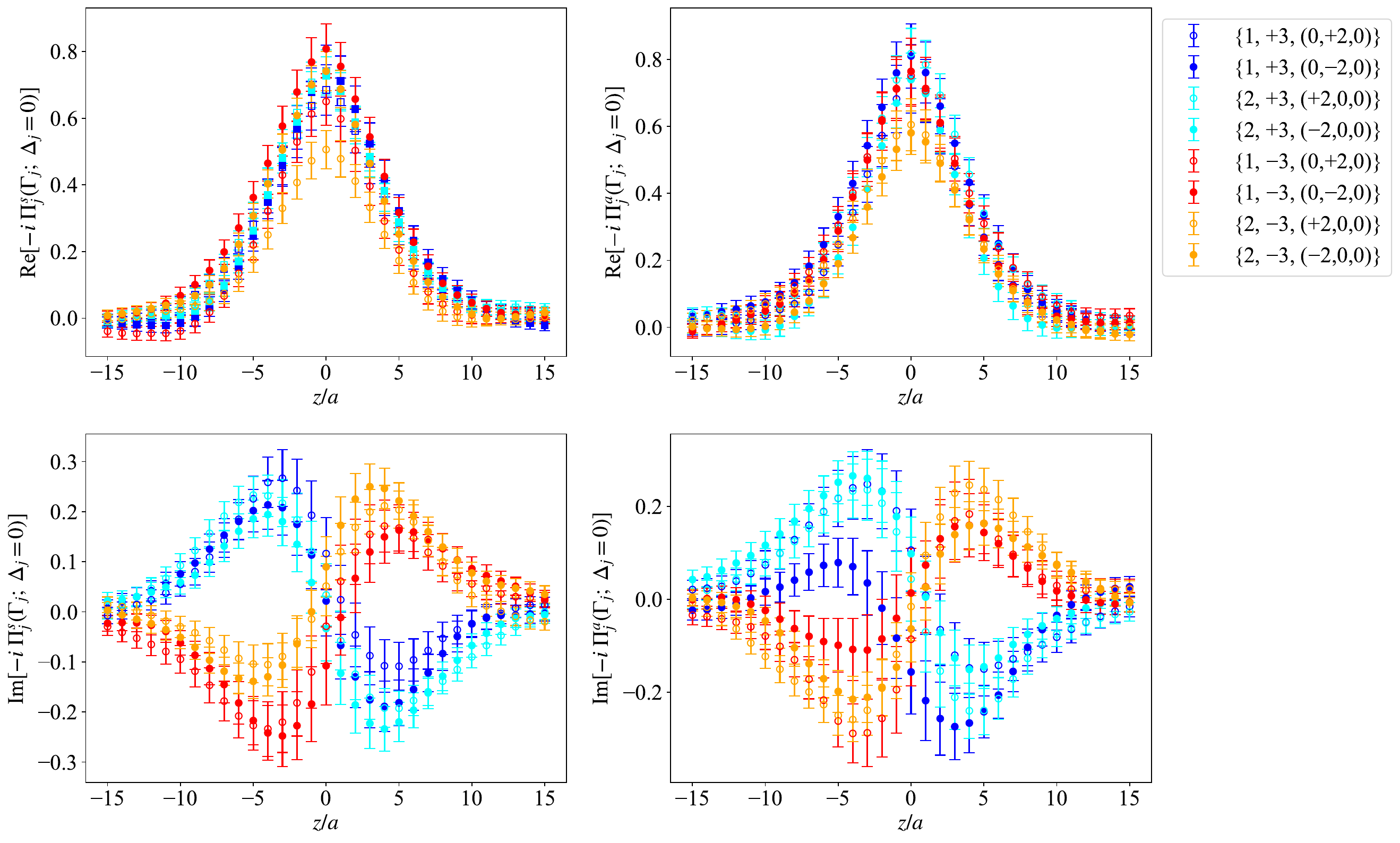}
    \vspace*{-0.35cm}
    \caption{\small Bare matrix elements $\Pi_1(\Gamma_1)$ and $\Pi_2(\Gamma_2)$ in the symmetric frame (left) and in the asymmetric frame (right). The legend indicates $\{j, P_3,\vec{\Delta}\}$ in units of $2\pi/L$, with $j$ corresponding to $\Pi_j(\Gamma_j)$. The remaining notation is the same as Fig.~\ref{fig:Pi3G3}.  }
    \label{fig:PijGj_A}
\end{figure}

\begin{figure}[h!]
    \centering
    \includegraphics[scale=0.38]{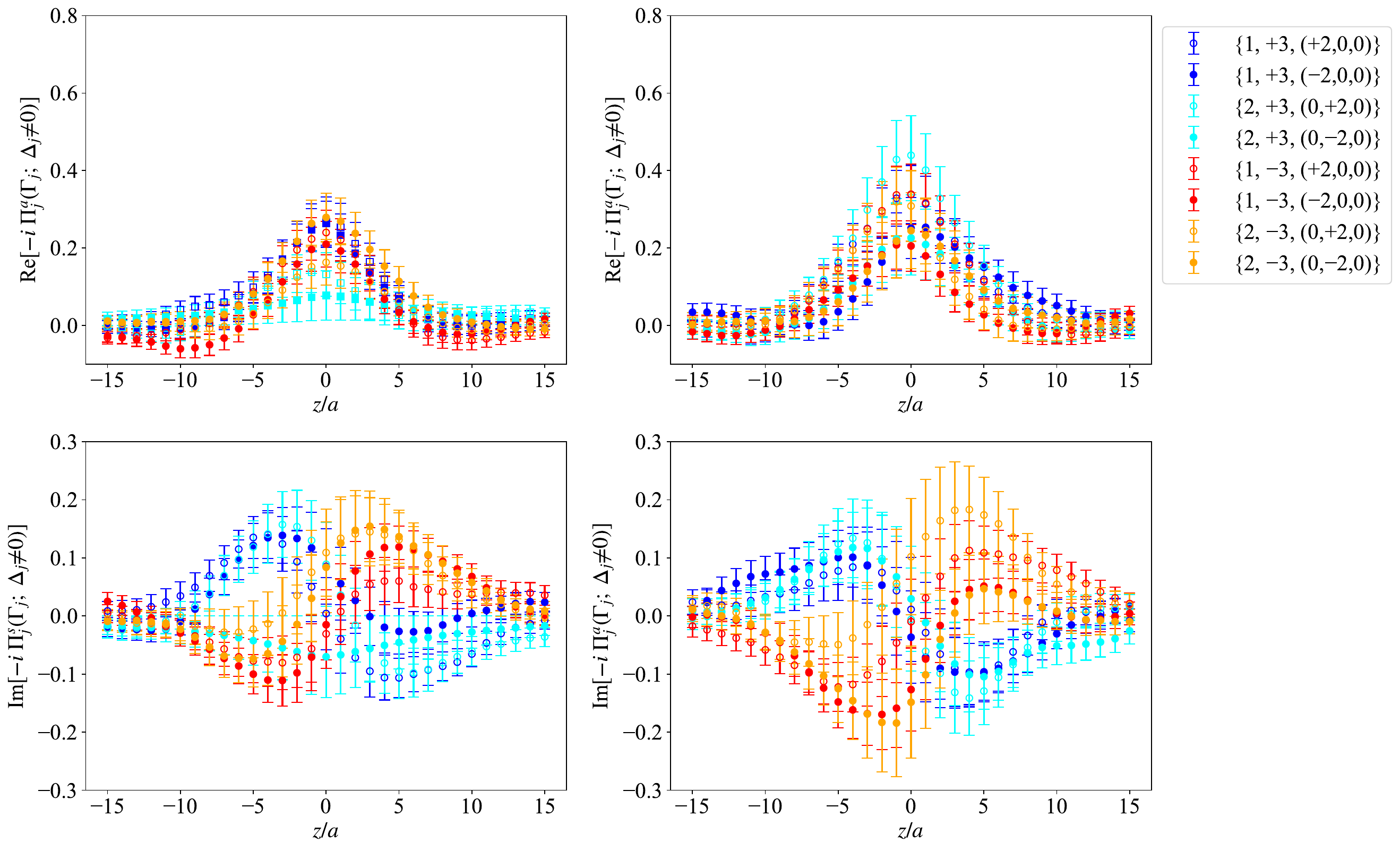}
    \vspace*{-0.35cm}
    \caption{\small Bare matrix elements $\Pi_1(\Gamma_1)$ and $\Pi_2(\Gamma_2)$ in the symmetric frame (left) and in the asymmetric frame (right). The legend indicates $\{j, P_3,\vec{\Delta}\}$ in units of $2\pi/L$, with $j$ corresponding to $\Pi_j(\Gamma_j)$. The remaining notation is the same as Fig.~\ref{fig:Pi3G3}.}
    \label{fig:PijGj_B}
\end{figure}

Below we present selected matrix elements in the two frames. 
Although direct numerical comparisons of these values may not yield direct physical insights, they do prove invaluable in assessing the signal quality and the extent of symmetry breaking concerning the sign of $P_3$ and $z$.
In Fig.~\ref{fig:Pi3G3}, we show the real and imaginary parts of the bare matrix element $\Pi_3(\Gamma_3)$ for the eight combinations of $a P_3=\pm \frac{2\pi}{L}3$ and $a \vec{\Delta}=\frac{2\pi}{L}(\pm 2,0,0),\,\frac{2\pi}{L}(0,\pm 2,0)$.
Similarly,  Figs.~\ref{fig:Pi0G3} - \ref{fig:PijGj_B} show $\Pi_0(\Gamma_0)$, $\Pi_j(\Gamma_j;\Delta_j=0)$, and $\Pi_j(\Gamma_j;\Delta_j\neq0)$ ($j=1,\,2$), respectively. 
Note that $\Pi_j(\Gamma_j)$ leads to independent equations for $\Delta_j=0$ and $\Delta_j\neq0$ (see, e.g.,  Eq.~\eqref{eq:Pi1G1_s}).
All the plots presented in this section offer side-by-side comparisons between the symmetric and asymmetric frame data. 
It is essential to underline that the numerical values of matrix elements in these two frames should not be directly compared. 
This is because the parametrization of a matrix element for a given operator and parity projector differs between frames in terms of the involved $\widetilde{A}_i$ and their associated kinematic coefficients. 
For example, $\Pi^s_3(\Gamma_3)$ contains information on $\widetilde{A}_2$, $\widetilde{A}_6$, and $\widetilde{A}_7$, while $\Pi^a_3(\Gamma_3)$ decomposes into $\widetilde{A}_2$, $\widetilde{A}_3$, $\widetilde{A}_4$, $\widetilde{A}_6$, and $\widetilde{A}_7$, as can be seen in Eqs.~\eqref{eq:Pi3G3_s}, \eqref{eq:Pi3G3_ns}.

Comparison of $\Pi_3(\Gamma_3)$ from Fig.~\ref{fig:Pi3G3} in the two frames reveals two features: (a) the matrix elements in the symmetric frame are less noisy than in the asymmetric frame; and (b) the variation of the data between the eight different classes of $\pm P_3$ and $\pm z$ is smaller in the symmetric frame. 
The latter is due to the fact that symmetries in the matrix elements with respect to $\pm P_3$ and $\pm z$ are only present in the symmetric frame.
Nevertheless, the asymmetry in $\pm P_3$ and $\pm z$ is found to be small for this kinematic setup in the asymmetric frame.
Similar observations hold for $\Pi_0(\Gamma_0)$ and $\Pi_{1,2}(\Gamma_{1,2})$, which are not shown here. 
In the light cone limit, the operators $\gamma_0\gamma_5$ and $\gamma_3\gamma_5$ are the components of $\gamma_+ \gamma_5$, and, thus, lead to the standard helicity GPDs, $\widetilde{H}$ and $\widetilde{E}$. 
This justifies the large magnitude observed in Figs.~\ref{fig:Pi3G3} - \ref{fig:Pi0G3}. 
However, $\gamma_0\gamma_5$ has finite mixing under renormalization, while $\gamma_3\gamma_5$ does not~\cite{Constantinou:2017sej}. The mixing was previously investigated numerically for twisted mass fermions~\cite{Constantinou:2017sej, Alexandrou:2017huk} and was found that the inclusion of a clover term in the fermion action suppresses the mixing significantly.

To extract the amplitudes $\widetilde{A}_i$, we utilize the matrix elements obtained from all possible combinations of operators and projectors, as detailed in Section~\ref{sec:mat_meth}. 
For each $\widetilde{A}_i$, we combine the positive and negative values of $P_3$, $\vec{\Delta}$, and $z$ according to their respective symmetry properties, as outlined in Eqs.~\eqref{e:HTR_cons}.
Upon averaging the data for $\widetilde{A}_i$, we proceed to compare their estimated values in the two frames. This comparison serves as a numerical assessment of their agreement, essentially acting as a consistency check for the lattice estimates of $\widetilde{A}_i$. The degree of agreement observed between the two frames offers an assessment of systematic effects, such as those stemming from finite lattice spacing, that may affect the results.

An essential aspect of our analysis concerns the treatment of $\widetilde{A}_3$, $\widetilde{A}_4$, and $\widetilde{A}_8$.
in Section~\ref{s:decomposition}, we provided theoretical arguments explaining why $\widetilde{A}_3$, $\widetilde{A}_4$, and $\widetilde{A}_8$ should vanish when the skewness is zero. 
For completeness, in Sec.~\ref{sec:mat_meth}, we presented the relations between the matrix elements and the $\widetilde{A}_i$, including all the amplitudes. 
In the symmetric frame, it is noteworthy that $\widetilde{A}_3$, $\widetilde{A}_4$, and $\widetilde{A}_8$ do not emerge in matrix elements associated with any of the remaining five $\widetilde{A}_i$ as they are exclusively linked to $\Pi_0^s(\Gamma_j)$, $\Pi_j^s(\Gamma_3)$, and $\Pi_3^s(\Gamma_j)$. 
Numerical exploration in the symmetric frame demonstrates that $\Pi_0^s(\Gamma_j)$, $\Pi_j^s(\Gamma_3)$, and $\Pi_3^s(\Gamma_j)$ are zero within uncertainties, validating that $\widetilde{A}_3$, $\widetilde{A}_4$, and $\widetilde{A}_8$ are indeed zero.
In contrast, the asymmetric frame exhibits a more intricate interplay among $\{\widetilde{A}_3,\,\widetilde{A}_4,\,\widetilde{A}_8\}$, and the remaining amplitudes.
Consequently, we perform an analysis of matrix elements, considering all $\widetilde{A}_i$ to numerically test whether the extracted $\widetilde{A}_3$, $\widetilde{A}_4$, and $\widetilde{A}_8$ are consistent with zero. 
Our findings confirm that $\widetilde{A}_3$, $\widetilde{A}_4$, and $\widetilde{A}_8$ are indeed zero within statistical errors in the asymmetric frame.
Furthermore, it is important to note that the numerical values of $\widetilde{A}^a_1$, $\widetilde{A}^a_2$, $\widetilde{A}^a_5$, $\widetilde{A}^a_6$, and $\widetilde{A}^a_7$ remain consistent, regardless of whether we include $\widetilde{A}_3$, $\widetilde{A}_4$, and $\widetilde{A}_8$ in the analysis or not. This consistency provides additional validation for the results obtained.

\begin{figure}[h!]
    \centering \includegraphics[scale=0.38]{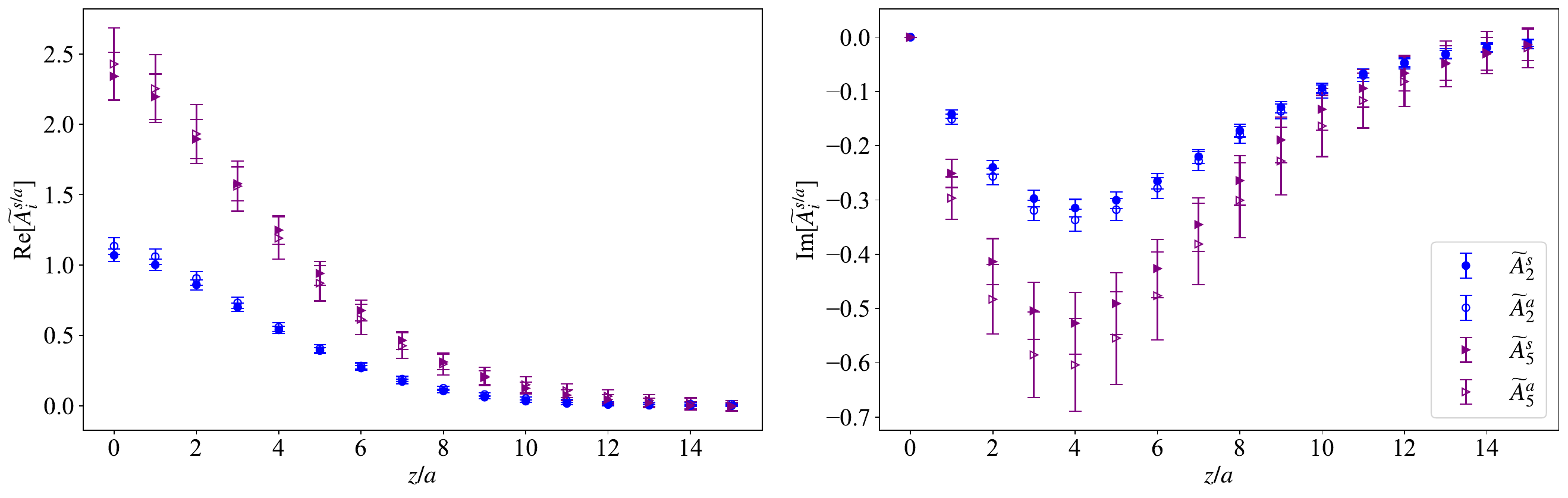}
    \vspace*{-0.3cm}
    \caption{\small Comparison of bare values of $\widetilde{A}_2$ and $\widetilde{A}_5$ in the symmetric (filled symbols) and asymmetric (open symbols) frame. The real (imaginary) part of each quantity is shown in the left (right) column. The data correspond to $|P_3|=1.25$ GeV and $-t=0.69$ GeV$^2$ ($-t=0.65$ GeV$^2$) for the symmetric (asymmetric) frame. }
    \label{fig:Ai_a}
\end{figure}

\begin{figure}[h!]
    \centering
\includegraphics[scale=0.38]{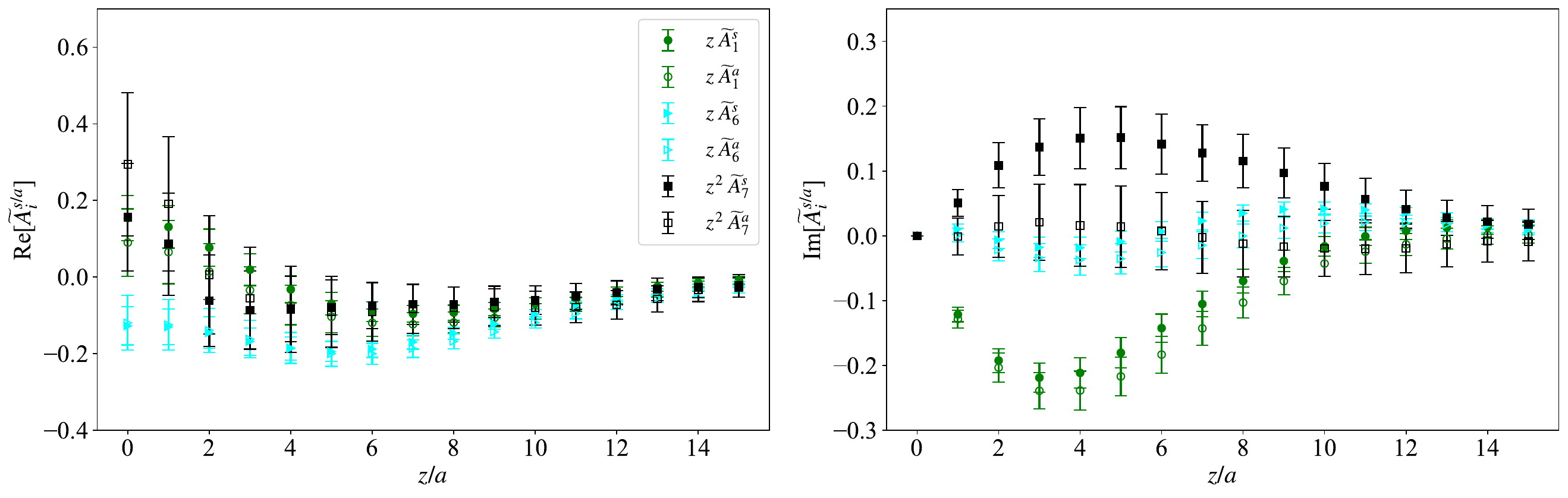}
    \vspace*{-0.3cm} 
    \caption{\small Comparison of bare values of $z\widetilde{A}_{1}$, $z\widetilde{A}_6$, and $z^2\widetilde{A}_7$ in the symmetric (filled symbols) and asymmetric (open symbols) frame. The notation is the same as Fig.~\ref{fig:Ai_a}.}
    \label{fig:Ai_b}
\end{figure}

In Figs.~\ref{fig:Ai_a} - \ref{fig:Ai_b} we present a comparison of the amplitudes using our data obtained with $P_3=\pm1.25$ GeV and $\vec{\Delta}=\{\frac{2\pi}{L}(\pm2,0,0), \frac{2\pi}{L}(0,\pm2,0)\}$ for both frames ($-t^s=0.69$ GeV$^2$ and $-t^a=0.65$ GeV$^2$). 
This comparison takes into account all eight combinations of $\pm P_3$ and $\pm\vec{\Delta}$. 
Among the amplitudes, we observe that $\widetilde{A}_5$ has the largest magnitude both in the real and the imaginary parts, followed by $\widetilde{A}_2$. 
The remaining amplitudes are notably small or negligible, which can be attributed to the small signal for certain matrix elements.
Encouragingly, we find a very good agreement between the two frames for each $\widetilde{A}_i$ up to statistical fluctiations, as expected given their Lorentz-invariant definition. 
As previously mentioned, the small differences observed may be associated with the approximately 5\% discrepancy between $t^s$ and $t^a$, as well as potential systematic uncertainties that have yet to be determined. 
Regarding the amplitudes $\widetilde{A}_1$, $\widetilde{A}_6$, and $\widetilde{A}_7$, they cannot be directly accessed at $z=0$ because their associated kinematic coefficients in Eqs.~\eqref{eq:Pi0G0_s} - \eqref{eq:Pi3G3_ns} become zero.
Nevertheless, one may perform extrapolations on their $z$ dependence to estimate $\widetilde{A}_i(z=0)$.
These findings collectively provide valuable insights into the behavior of the amplitudes under various conditions and kinematic setups.

\begin{figure}[h!]
    \centering
        \includegraphics[scale=0.38]{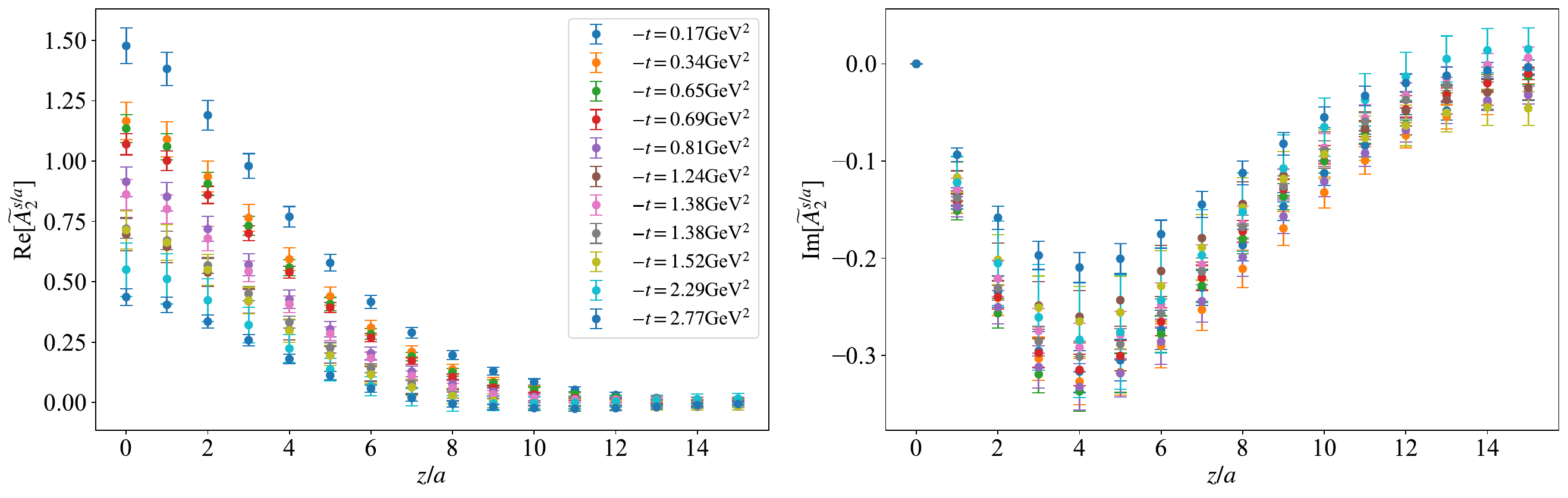}
    \vspace*{-0.3cm} 
    \caption{\small The amplitude $\widetilde{A}_2$ for all values of $-t$ given in Table~\ref{tab:stat}.}
    \label{fig:A2_t}
\end{figure}

\begin{figure}[h!]
    \centering
        \includegraphics[scale=0.38]{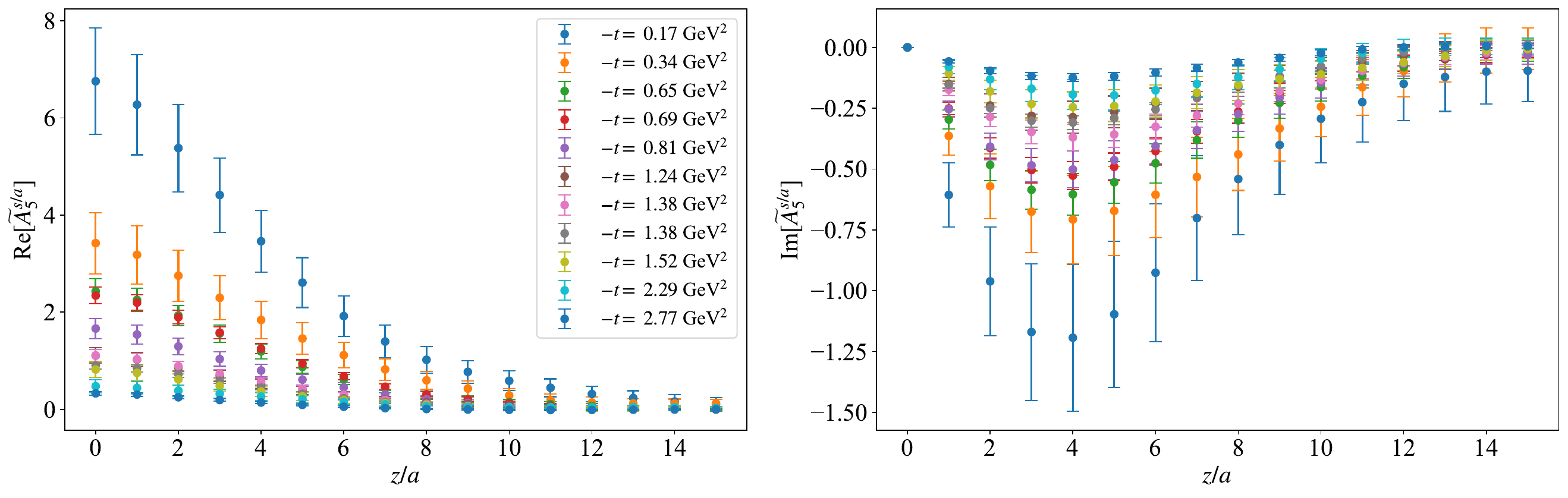}
    \vspace*{-0.3cm} 
    \caption{\small The amplitudes  $\widetilde{A}_5$ for all values of $-t$ given in Table~\ref{tab:stat}.}
    \label{fig:A5_t}
\end{figure}

An insightful exploration of the amplitudes involves examining their dependence on momentum transfer. 
To illustrate this, we will focus on the substantial amplitudes,  $\widetilde{A}_2$ and $\widetilde{A}_5$, and analyze their $t$ dependence. 
The results for these amplitudes are presented in Fig.~\ref{fig:A2_t} and Fig.~\ref{fig:A5_t}, respectively. 
We use all the data obtained in this work that covers the range $-t \in [0.17-2.77]$ GeV$^2$.
Importantly, since the amplitudes are frame invariant, a single function can describe the data from any frame. 
This feature allows for a direct comparison of the data, ensuring consistency in the analysis of the two frames.
Our observations reveal that as $-t$ increases, both the real and imaginary parts of the amplitudes decrease in magnitude.
It is noteworthy that, based on our findings, these amplitudes continue to exhibit non-zero values even at $-t$ beyond 2 GeV$^2$.
However, it is essential to exercise caution in this high-momentum transfer region. The calculations may suffer from systematic uncertainties and higher-twist contamination, rendering this region less reliable for precise conclusions.

It is worth noting that the presence of the pion pole discussed in Refs.~\cite{Bhattacharya:2022xxw,Bhattacharya:2023wvy} is argued to extend to the GPD $\widetilde{E}$ in the isovector channel, where $\widetilde{E}^u - \widetilde{E}^d \sim \tfrac{1}{t-m^{2}_\pi}$~\cite{Penttinen:1999th}. As discussed in Appendix~\ref{s:local}, the trace of this pole should be evident in $\widetilde{A}_5$.
By comparing the $t$-dependence of $\widetilde{A}_5$ with, for example, $\widetilde{A}_2$ around $z=0$, we can infer that the lattice results for $\widetilde{A}_5$ contains a pion pole behavior, and so should its Mellin moments, for example $g_P$. 
It is worth highlighting that this $z=0$ scenario aligns with our previously established anticipations for the pseudo-scalar form factor exhibiting a pole, as discussed in Ref.~\cite{Alexandrou:2020okk}. 
The significance of this finding is underscored by the considerably steeper rise of $\widetilde{A}_5$ compared to the other amplitudes.
For instance, as observed from the plots in this section, in the $t$ range of $[-0.65, -0.17], \textrm{GeV}^2$, the rise factor is approximately 2.7 for $\widetilde{A}_5$, while for $\widetilde{A}_2$ it is around 1.3. 
Similarly, in the $t$ range of $[-2.29, -0.17] , \textrm{GeV}^2$, the rise factor is roughly 15 for $\widetilde{A}_5$, whereas it is about 2.7 for $\widetilde{A}_2$. 
These findings collectively support the notion that the detection of this pole does not necessitate a full calculation of $\widetilde{E}$. 
In the near future, we intend to thoroughly investigate this phenomenon in conjunction with the examination of the manifestation of the pole within the $z$-dependence of the amplitude.
Such extension of our analysis aims to provide a deeper understanding of the pole's influence beyond the context of moments to a broader and more holistic understanding of its impact on the physics under consideration.

\subsection{Quasi-GPDs in coordinate space}

\begin{figure}[h!]
    \centering
\includegraphics[scale=0.38]{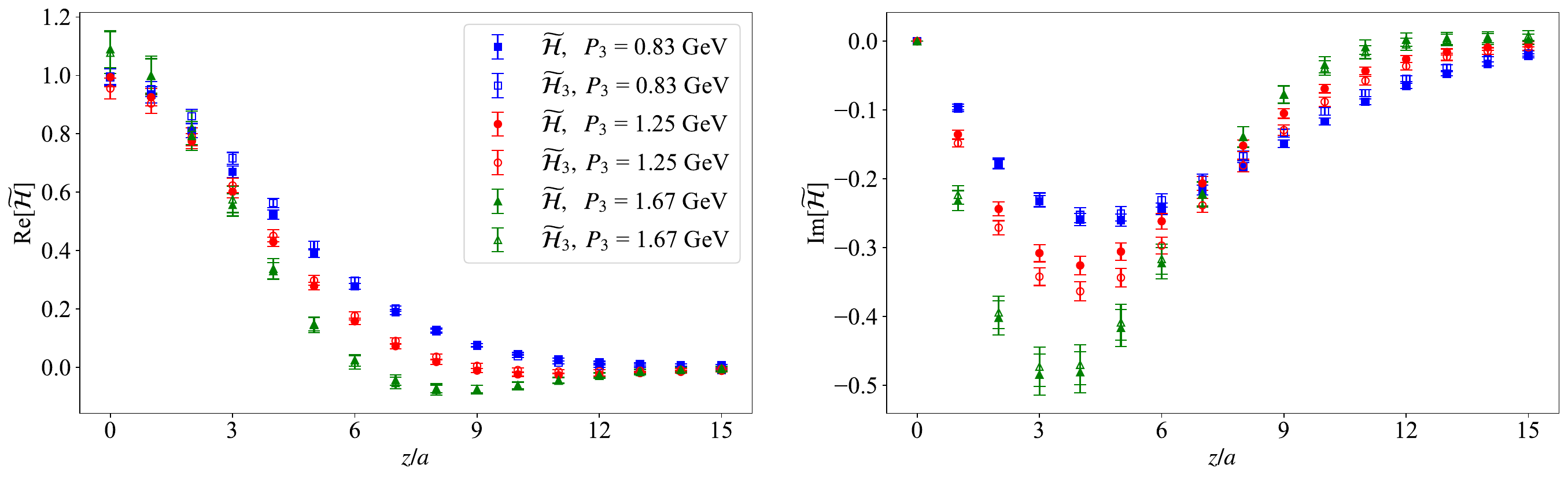}
    \vspace*{-0.5cm} 
    \caption{\small Momentum boost dependence for $\widetilde{\cal H}_3$ (Eq.~\eqref{eq:FH_s}) and $\widetilde{\cal H}$ (Eq.~\eqref{eq:FH_LI}) at $-t=0.69$ GeV$^2$.}
    \label{fig:FH_P3}
\end{figure}

\begin{figure}[h!]
    \centering
\includegraphics[scale=0.38]{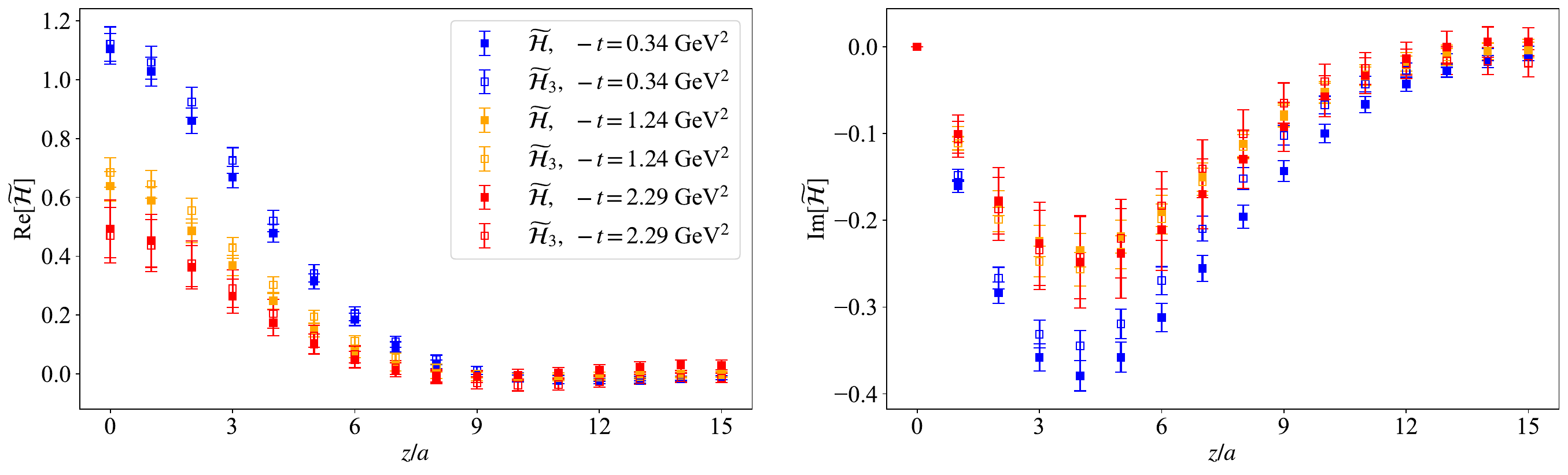}
    \vskip -0.5cm 
    \caption{\small Comparison of $\widetilde{\cal H}_3$ and $\widetilde{\cal H}$ using the asymmetric frame data with $-t=0.34,\,1.24,\,2.29$ GeV$^2$.}
    \label{fig:FH3_FH_a2}
\end{figure}

Our attention now shifts to the quasi-GPDs, which are renormalized in coordinate space using the RI$'$ prescription developed and refined in Refs.~\cite{Constantinou:2017sej, Alexandrou:2017huk, Alexandrou:2019lfo}.
We refer the reader to these publications for more details, as well as the previous work of Ref.~\cite{Bhattacharya:2022aob}.
In Sec.~\ref{s:decomposition}, we discussed that the definition of quasi-GPDs is not unique. 
In this context, we will consider the standard $\gamma_3\gamma_5$ definition, denoted as $\widetilde{\cal H}_3$, as well as an alternative definition that is constructed to be Lorentz invariant, termed $\widetilde{\cal H}$. 
As mentioned above, $\widetilde{\cal H}_3$ also exhibits frame independence, in contrast to the case of the unpolarized GPDs.
This frame independence is linked to the fact that the indices of the axial operator align with the direction of the momentum boost, a point we discussed in detail in Section~\ref{s:decomposition}. 
The relations between the quasi-$H$ GPD and the amplitudes are provided in Eq.~\eqref{eq:FH_s} and Eq.~\eqref{eq:FH_LI}, for $\widetilde{\cal H}_3$ and $\widetilde{\cal H}$, respectively.
To enable the comparison of momentum boost dependence, we utilize the data in the symmetric frame at $-t^s=0.69$ GeV$^2$ considering different values of the momentum boost $P_3=0.83,\,1.25,\,1.67$ GeV.
The comparison between $\widetilde{\cal H}_3$ and $\widetilde{\cal H}$ is presented for each value of $P_3$ in Fig~\ref{fig:FH_P3}.
This comparison provides valuable insights into the behavior of the quasi-GPDs in relation to the different definition.
Interestingly, the two definitions of the quasi-GPD are compatible for all cases.
Although there is a slight difference in the imaginary part for intermediate values of $z$ at $P_3=1.25$ GeV, overall, the two definitions remain consistent.
In terms of the $P_3$ dependence, we find that, as $P_3$ increases, the real part approaches zero at smaller values of $z$. 
On the other hand, the imaginary part is enhanced for higher $P_3$, a feature also observed in PDFs.
This trend suggests a strong link between the behavior of quasi-GPDs and PDFs in response to varying value for the momentum boost.
\begin{figure}[h!]
    \centering
\includegraphics[scale=0.38]{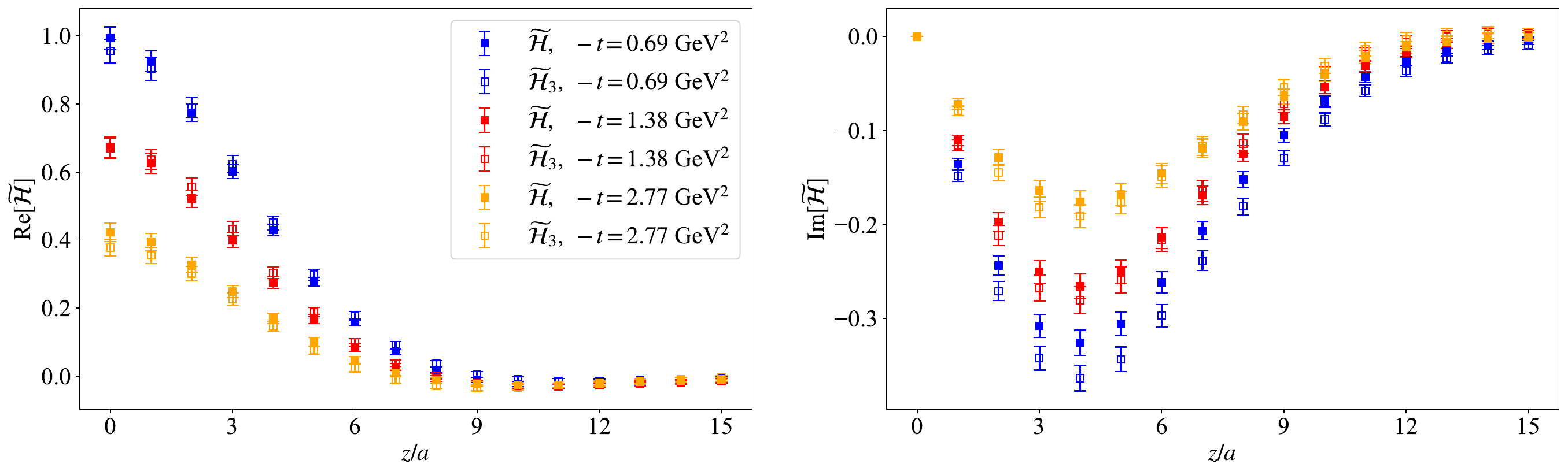}
\vspace*{-0.5cm} 
\caption{\small Comparison of $\widetilde{\cal H}_3$ and $\widetilde{\cal H}$ using the symmetric frame data with $-t=0.69,\,1.38,\,2.77$ GeV$^2$.}
    \label{fig:FH3_FH_s}
\end{figure}
\begin{figure}[h!]
    \centering
\includegraphics[scale=0.38]{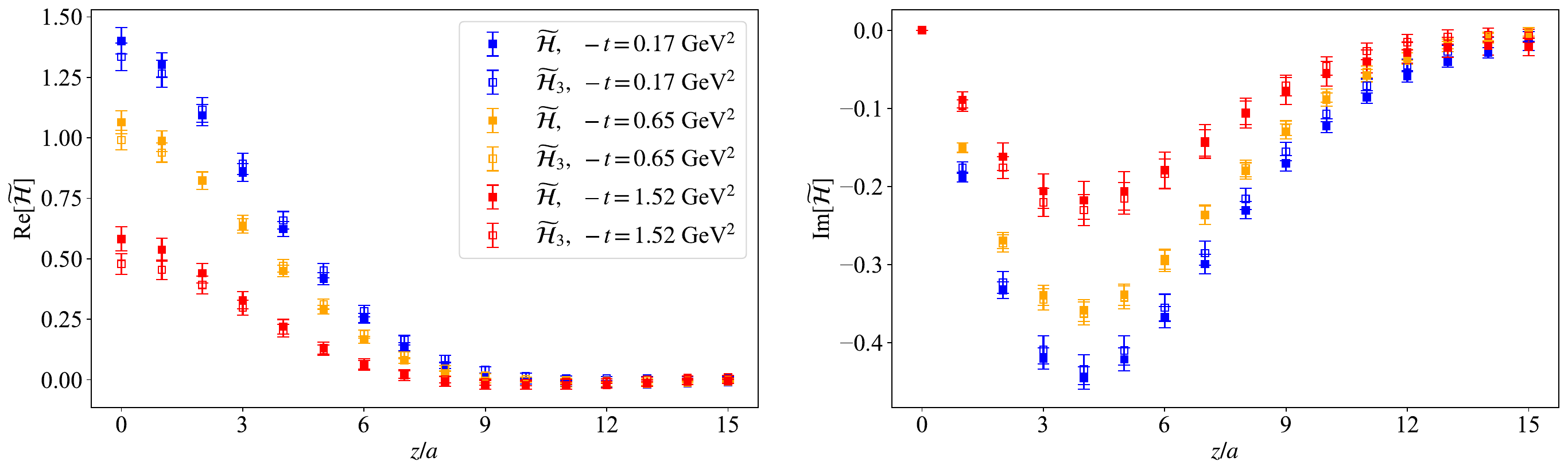}
    \vspace*{-0.5cm} 
    \caption{\small Comparison of $\widetilde{\cal H}_3$ and $\widetilde{\cal H}$ using the asymmetric frame data with $-t=0.17,\,0.65,\,1.52$ GeV$^2$. }
    \label{fig:FH3_FH_a1}
\end{figure}
To investigate whether the similarity between the two definitions of quasi-$\widetilde{H}$ GPD is specific to the particular $-t=0.69$ GeV$^2$, we extend the analysis to other values of $t$. 
We compare the two definitions for all the data in both the symmetric frame (Fig.~\ref{fig:FH3_FH_s}) and the asymmetric frame (Figs.~\ref{fig:FH3_FH_a1} - \ref{fig:FH3_FH_a2}).
Remarkably, we consistently find agreement between the two definitions across both frames. This agreement is noteworthy because, theoretically, quasi-GPD definitions are not unique, and one might expect variations in the results. The level of agreement observed suggests a robustness in the analysis and interpretation of these GPDs, despite the lack of a unique definition.

\begin{figure}[h!]
    \centering
\includegraphics[scale=0.38]{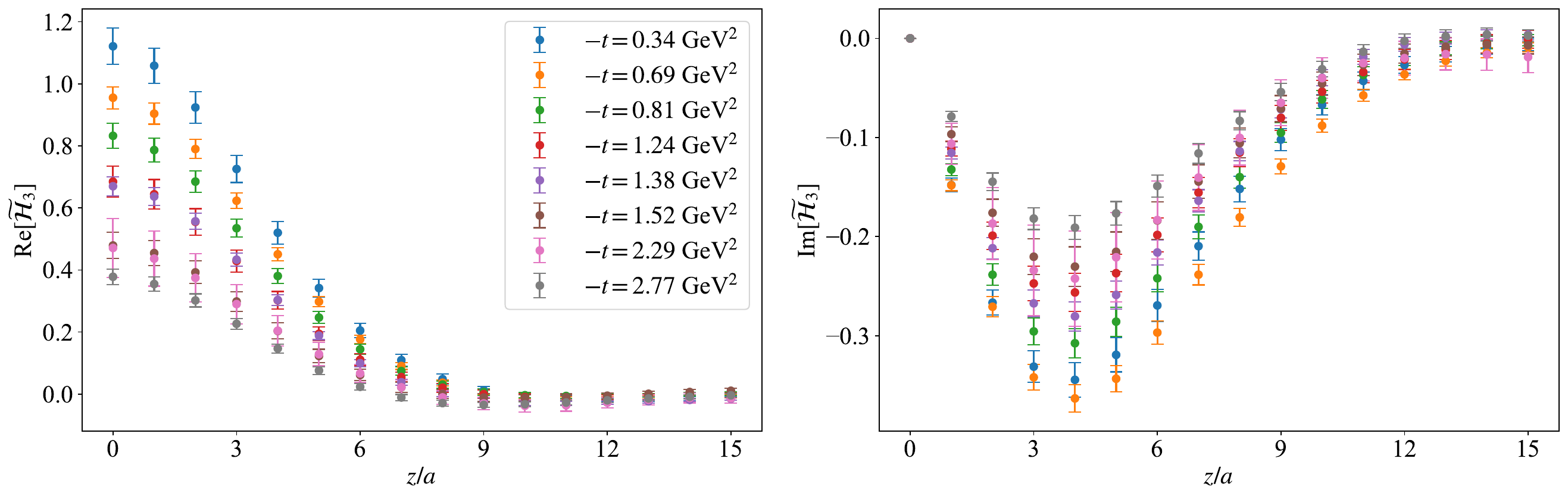}
    \vspace*{-0.5cm} 
    \caption{\small The quasi-GPD $\widetilde{\cal H}_3$ for several values of the momentum transfer squared, $-t$ in coordinate space.}
    \label{fig:FH3_t}
\end{figure}

\begin{figure}[h!]
    \centering
\includegraphics[scale=0.4]{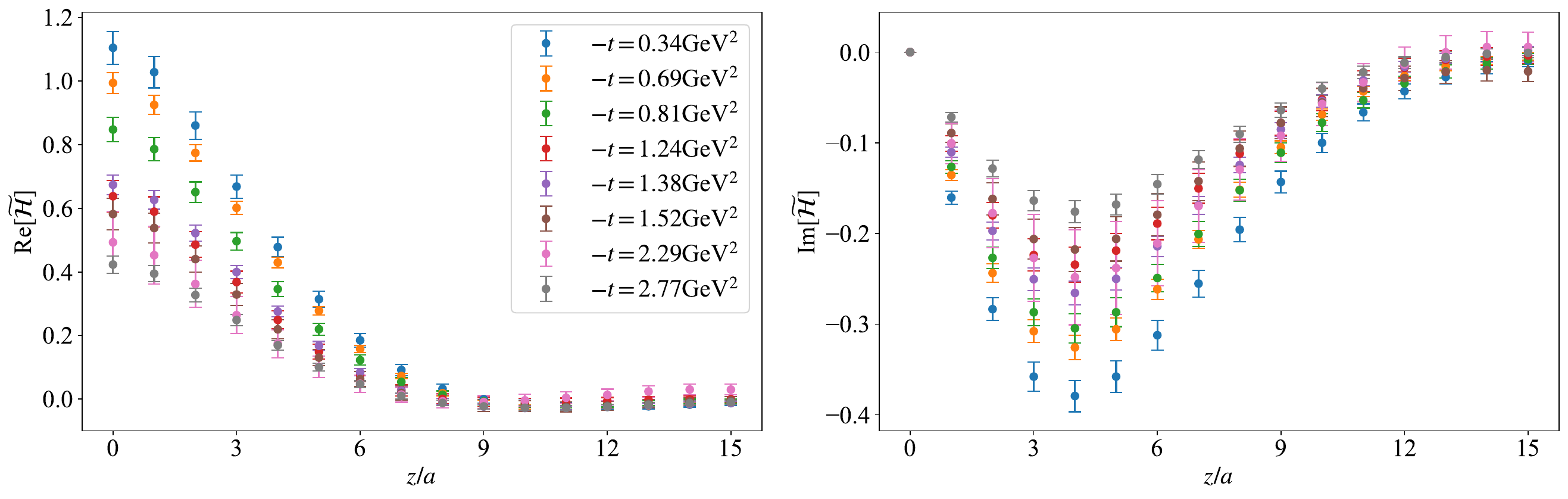}
    \vspace*{-0.5cm} 
    \caption{\small The quasi-GPD $\widetilde{\cal H}$ for various values of the momentum transfer squared, $-t$ in coordinate space.}
    \label{fig:FH_t}
\end{figure}

In Figs.~\ref{fig:FH3_t} - \ref{fig:FH_t}, we present the $-t$ dependence of $\widetilde{\cal H}_3$ and $\widetilde{\cal H}$ in coordinate space to provide a comprehensive overview. 
The shapes of these functions exhibit striking similarities, as previously discussed.
However, there are notable distinctions worth highlighting.
For the real part, at $-t=0.69 {\rm GeV}^2$, $\widetilde{\cal H}_3$ maintains an equidistant relationship from both $\widetilde{\cal H}_3(-t=0.34 {\rm GeV}^2)$ and $\widetilde{\cal H}_3(-t=0.81{\rm GeV}^2)$. 
In contrast, $\widetilde{\cal H}(-t=0.69{\rm GeV}^2)$ shifts closer to $\widetilde{\cal H}_3(-t=0.34{\rm GeV}^2)$. 
The difference between the pair ${\widetilde{\cal H}_3(-t=1.24{\rm GeV}^2), \widetilde{\cal H}_3(-t=1.38{\rm GeV}^2)}$ and $\widetilde{\cal H}_3(-t=1.52{\rm GeV}^2)$ is more pronounced compared to the analogous comparison in the $\widetilde{\cal H}$ definition.
For the imaginary part, $\widetilde{\cal H}_3(-t=0.34{\rm GeV}^2)$ demonstrates compatibility with $\widetilde{\cal H}_3(-t=0.69{\rm GeV}^2)$. In the case of $\widetilde{\cal H}$, there is a discernible difference.


\subsection{Light-cone GPDs}

\begin{figure}[h!]
    \centering
\includegraphics[scale=0.38]{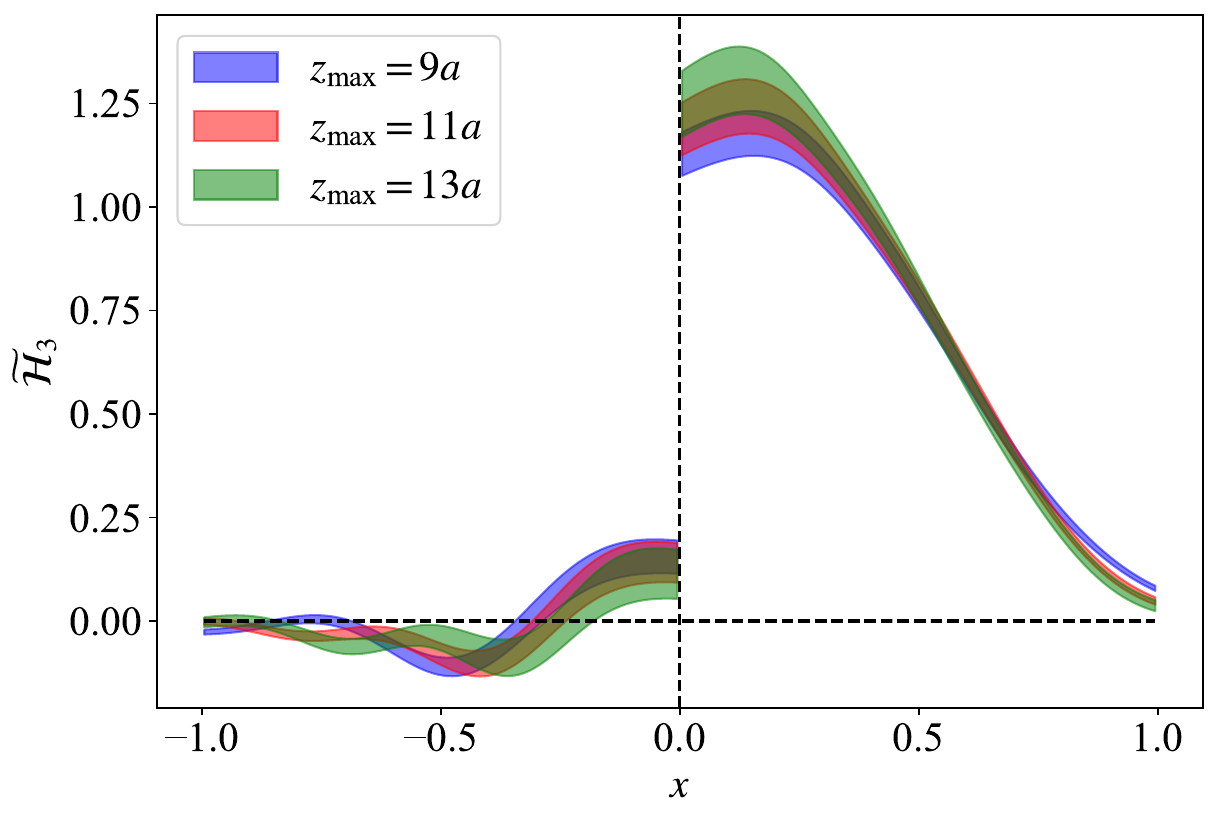}\quad
\includegraphics[scale=0.38]{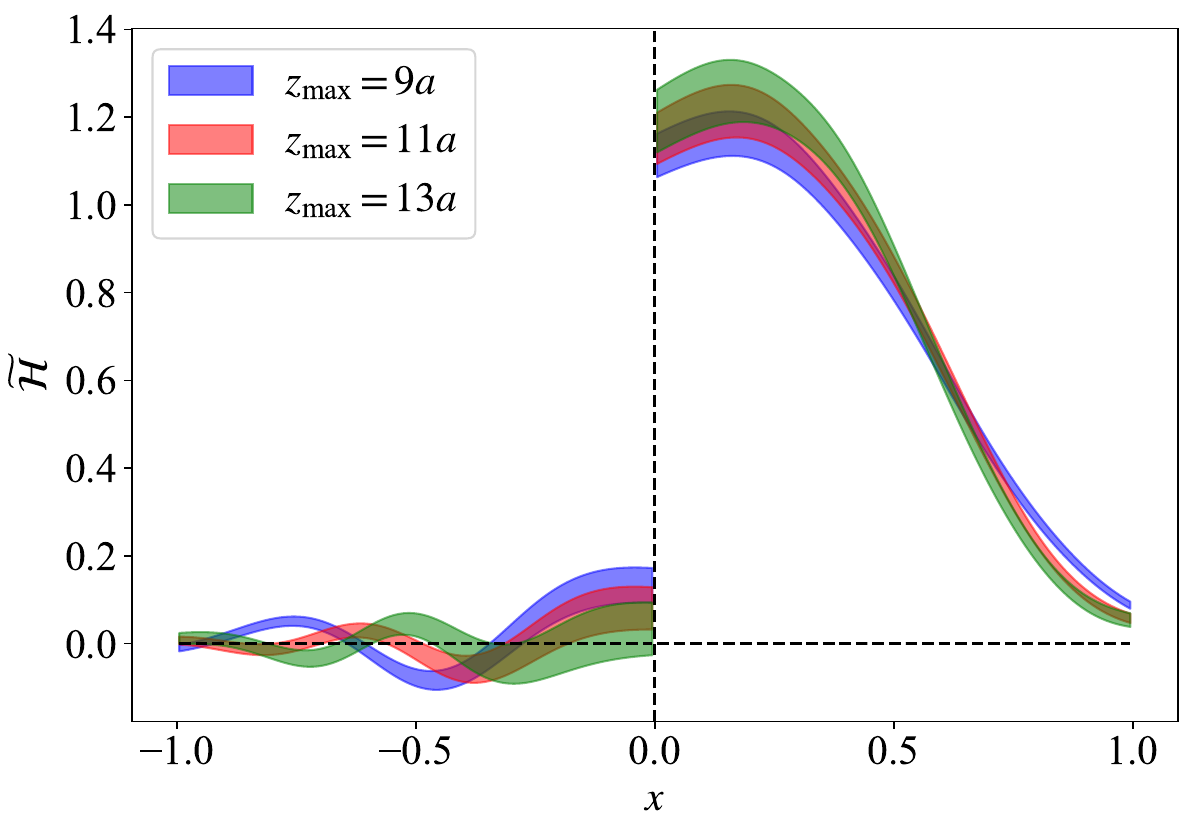}
    \vspace*{-0.5cm} 
    \caption{\small The $x$-dependent quasi-GPD $\widetilde{\cal H}_3$ (left) and $\widetilde{\cal H}$ (right) for $t^a=0.65$ GeV$^2$ and $P_3=1.25$ GeV using Backus-Gilbert with $z_{\rm max}=9a,\,11a,\,13a$.}
    \label{fig:FH3_H_x_zmax}
\end{figure}

The reconstruction of the $x$ dependence of the quasi-GPDs is not unique for a finite number of discrete data, as the standard Fourier transform suffers from the so-called inverse problem\footnote{See Ref.~\cite{Karpie:2019eiq} for an extensive discussion in the context of reconstructing partonic distributions.}, which intensifies in the small-$x$ region.
In this work, we employ the Backus-Gilbert (BG) reconstruction method~\cite{BackusGilbert}, which offers a solution to the inverse problem. This approach is based on a model-independent criterion to choose the light-cone reconstructed GPDs from the infinite set of possible solutions to the inverse problem. 
The criterion employed is that the variance of the solution with respect to the statistical variation of the input data should be minimal. 
Despite the model independence of the Backus-Gilbert method, its reliable applicability may be limited by the small number of lattice data sets that enter the reconstruction. 
An approach to assess the consistency and reliability of our reconstruction is to conduct a sensitivity test by varying the number of input data used in the Backus-Gilbert method.
Specifically, we test for different values of $z_{\rm max}$, namely $z_{\rm max}=9a,\,11a,\,13a$. 
Furthermore, the functions $\widetilde{\cal H}_3(z)$ and $\widetilde{\cal H}(z)$ for $z>z_{\rm max}$ are assumed to be zero.
This assumption helps establish a condition for the reconstruction and is a practical choice for such cases where data beyond this limit are either unavailable or less reliable.
Fig.~\ref{fig:FH3_H_x_zmax} shows the $x$ dependence of the quasi-GPD for the three values of $z_{\rm max}$. It is found that, for all $z_{\rm max}$ values, there is compatibility for the $\widetilde{\cal{H}}$ quasi-GPD up to $x = 0.7$, as well as consistency between $z_{\rm max}=11a$ and $z_{\rm max}=13a$ up to $x=1$. With that, $z_{\rm max}=11a$ has been chosen for the quasi-GPD to proceed with the matching to the light-cone GPDs. For the latter, we use the one-loop equations of Ref.~\cite{Liu:2019urm} at $\xi \rightarrow 0$, which we include here for completeness.
\begin{equation}
    \label{eq:matching}
    q(x) = \int^{+\infty}_{-\infty} dy~f_1\left(\Gamma,y,\xi=0,\frac{p^z}{\mu}\right)_+ \tilde{q}(y) + O \bigg (\frac{M^2}{P_3^2}, \frac{t}{P_3^2}, \frac{\Lambda_{\rm QCD}^2}{x^2 (1-x) P_3^2} \bigg )\,,
\end{equation}
\begin{equation}
    \label{eq:kernel}
f_1\left(\Gamma,y,\xi=0,\frac{p^z}{\mu}\right) = \frac{\alpha_sC_F}{2\pi} \left\{
        \begin{array}{ll}
            \frac{y^2+1}{x-1}\ln{\left(\frac{y}{y-1}\right)} - 1 & \quad  y< 0 \,, \\
            \frac{1+y^2}{1-y}\left[\ln{\frac{4y(1-y)(p^z)^2}{\mu^2}-1}\right]-2y+3 & \quad 0< y < 1 \,,\\
            -\frac{y^2+1}{y-1}\ln{\left(\frac{y}{y-1}\right)} + 1 & \quad y > 1\,.
        \end{array}
    \right.
\end{equation}
In the above equations, $\tilde{q}$ denoted a general quasi-GPD, $q$ is the corresponding light-cone GPD, and $f_1$ is the matching kernel. Since quasi-GPDs are defined in the RI scheme, the kernel contains the so-called RI counterterms in addition to $f_1$. The expressions are lengthy and can be found in Ref.~\cite{LatticeParton:2018gjr}. The matching formalism is constructed with the quasi-GPDs being defined in RI scheme at a scale of 1.2 GeV, while the light-cone GPDs are in the $\overline{\rm MS}$ scheme at a renormalization scale of 2 GeV. 

\begin{figure}[h!]
    \centering
\includegraphics[scale=0.4]{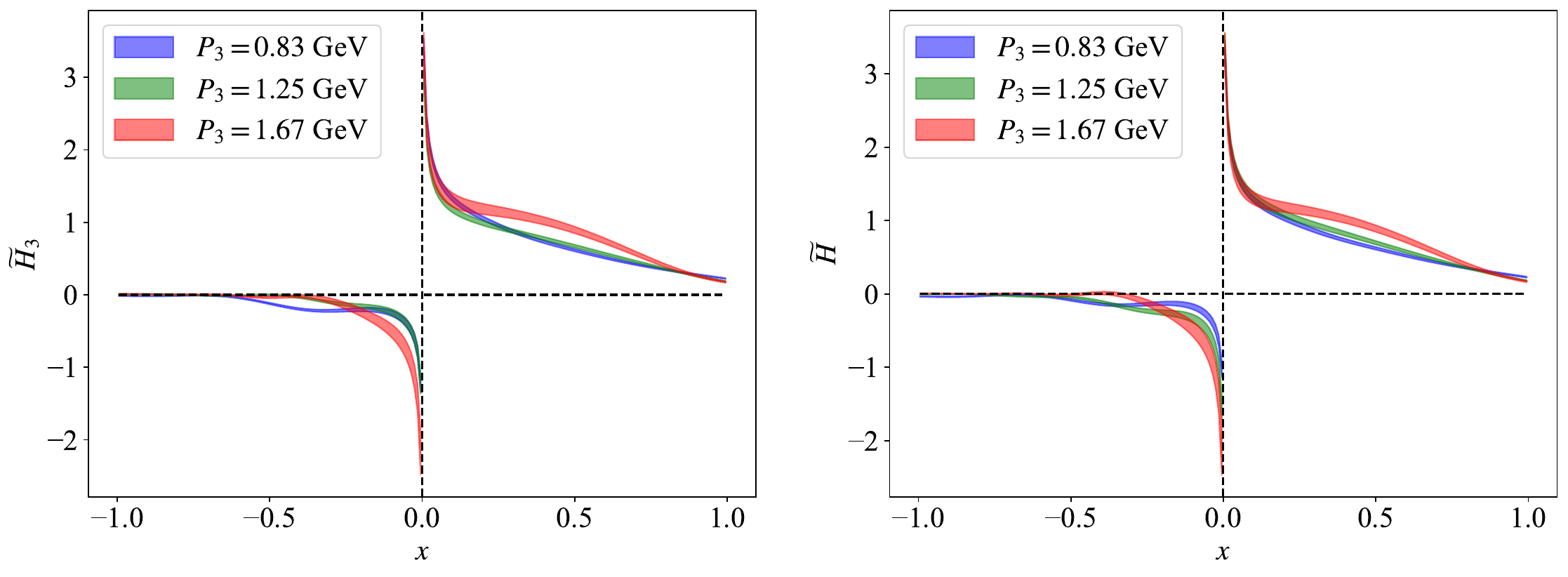}\quad
    \vspace*{-0.3cm} 
    \caption{\small The light-cone GPD $\widetilde{H}_3$ (left) and $\widetilde{H}$ (right) at $-t=0.69$ GeV$^2$ and for three values of the momentum boost, $P_3=0.83,\,1.25,\,1.67$ GeV. Results are given in the $\MSb$ scheme at 2 GeV.}
    \label{fig:matched_P_dependence}
\end{figure}

\begin{figure}[h!]
    \centering
\includegraphics[scale=0.278]{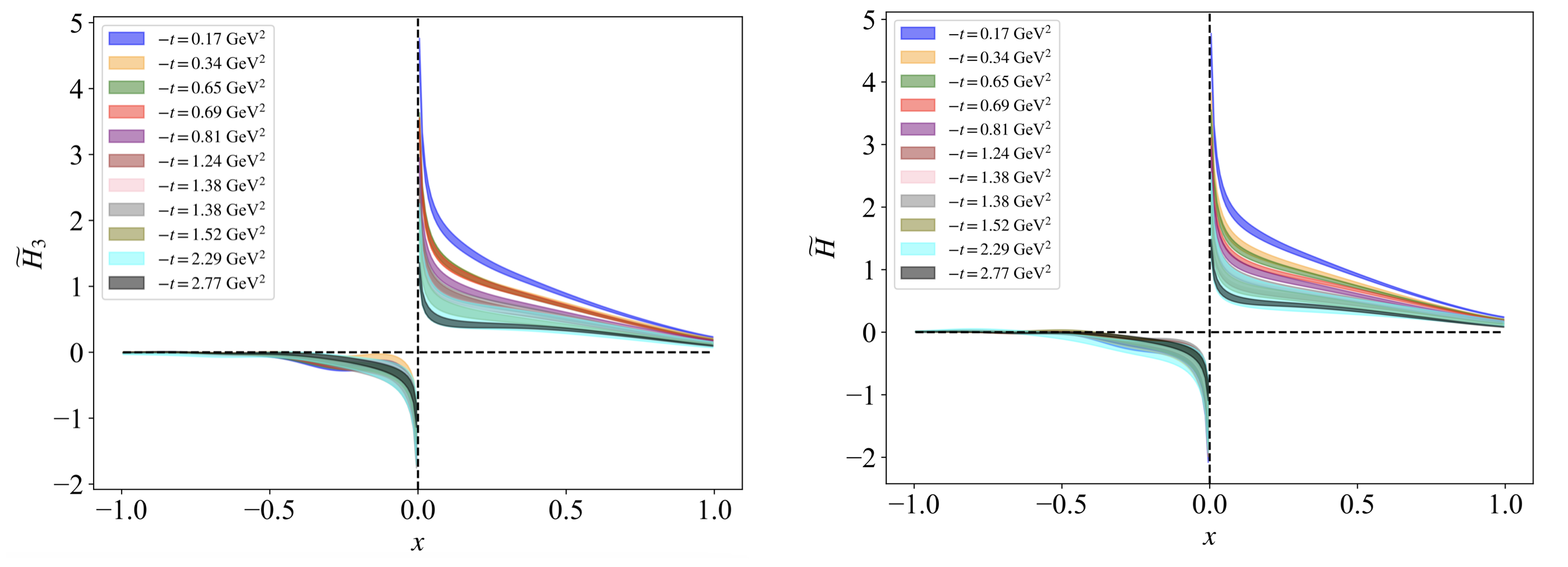}\quad
    \vspace*{-0.3cm} 
    \caption{\small The momentum-transfer squared dependence of the light-cone GPD $\widetilde{H}_3$ (left) and $\widetilde{H}$ (right) at $|P_3| = 1.25~\rm{GeV}$. Results are given in the $\MSb$ scheme at 2 GeV.}
    \label{fig:t_dependence_GPD}
\end{figure}

We have conducted a series of investigations for quasi-GPDs, which can also be performed for the light-cone cases. 
In this section, we provide a comparison between the two definitions for $\widetilde{H}$, examining their $P_3$ dependence where data is available, as well as their behavior with respect to $-t$. 
Fig.~\ref{fig:matched_P_dependence} illustrates the momentum boost dependence for both definitions, $\widetilde{H}_3$ and $\widetilde{H}$, specifically for the symmetric-frame data at $-t=0.69$ GeV$^2$. We observe some residual dependence between the two, albeit it is important to note that the difference between $P_3=1.25\rm{GeV}$ and $P_3=1.67\rm{GeV}$ is expected to be well within unquantified systematic uncertainties.
The behavior of the two definitions for $H$ GPD is remarkably similar, with only minor distinctions between $P_3=1.25\rm{GeV}$ and $P_3=1.67\rm{GeV}$. 
It is worth emphasizing that although the two definitions are different, they both possess Lorentz invariance. 
Consequently, each definition offers a unique function that is applicable in any frame. 
Turning to Fig.~\ref{fig:t_dependence_GPD}, we shift our focus to the $-t$ dependence of $\widetilde{H}_3$ and $\widetilde{H}$ at $|P_3|=1.25~\rm{GeV}$, where we have a substantial amount of data. 
Numerical values and statistical uncertainties are found to be similar for both definitions. 
This result is somewhat expected, as the difference between the two definitions is proportionally tied to $A_7$, which is found to be very small.

The GPDs exhibit a decaying behavior as $-t$ increases, which parallels the behavior of the form factors. 
Notably, for $-t>1.5~\rm{GeV}^2$, we observe a negligible dependence on $-t$, where the GPDs become of similar magnitude. 
It is important to recognize that this observation is qualitative in nature, as at such values of $-t$ and for $P_3=1.25$ GeV, the lattice results have increased higher-twist contamination. 
Nevertheless, we have included this data, as it was obtained at no additional computational cost due to the use of an asymmetric kinematic frame.
To conclude this discussion, we provide the data of Fig.~\ref{fig:t_dependence_GPD} in a three-dimensional plot to demonstrate both the $-t$ and $x$ dependence of $H$ GPD. 
For completeness, we show both definitions we explored in this work.

\begin{figure}[h!]
    \centering
\includegraphics[scale=0.33]{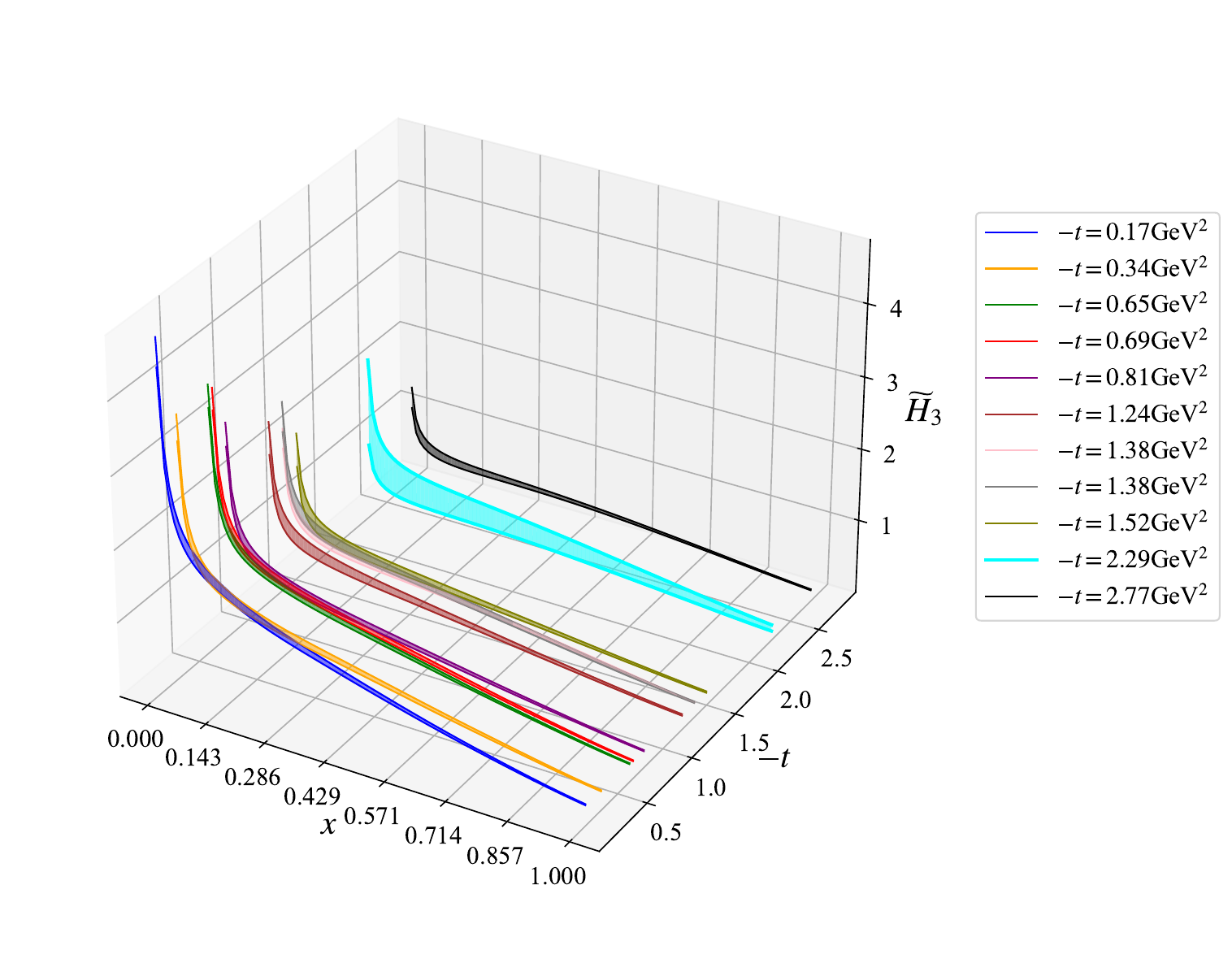}
\includegraphics[scale=0.33]{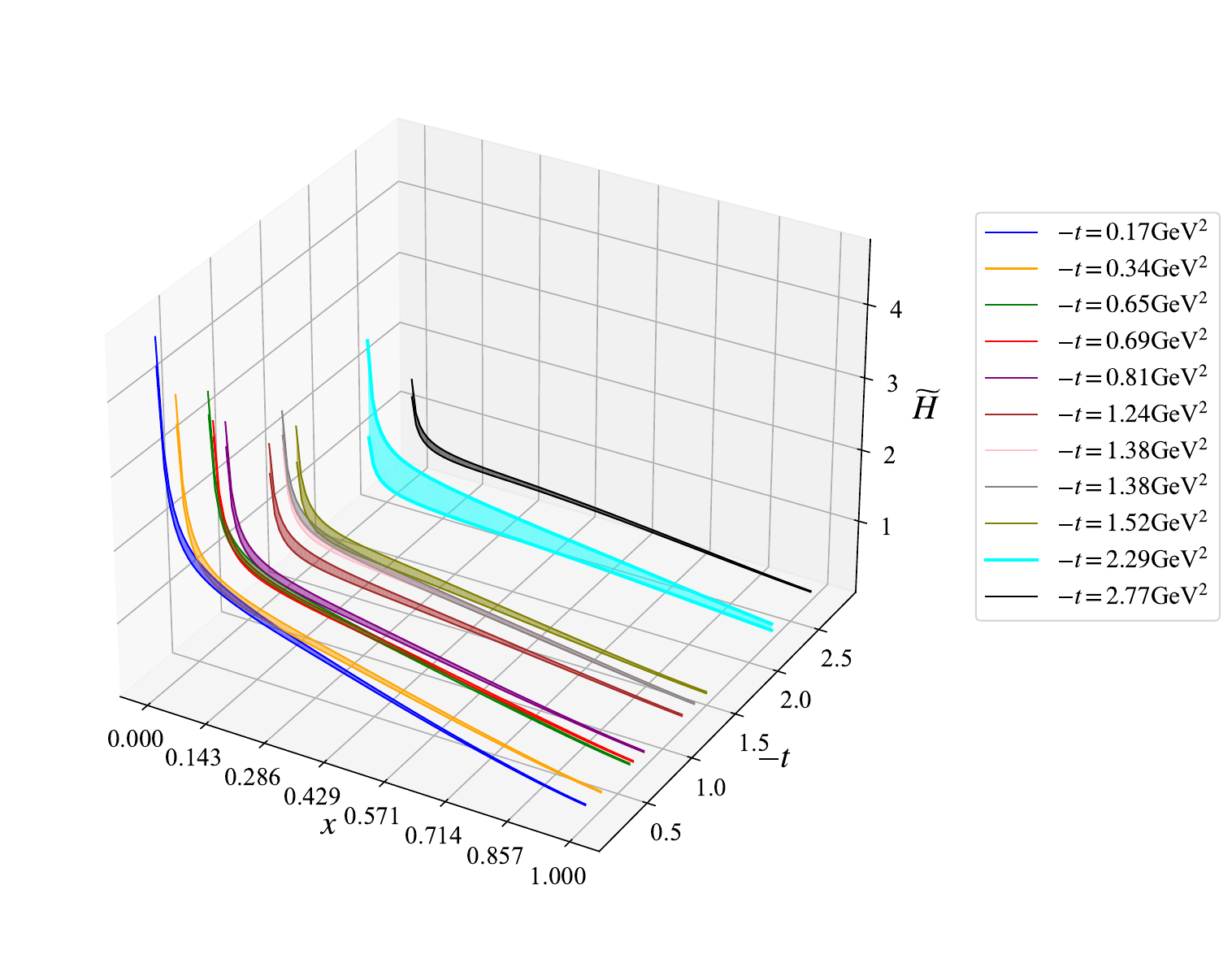}
    \vspace*{-0.3cm} 
    \caption{\small The $-t$ and $x$ dependence of the light-cone GPD $\widetilde{H}_3$ (left) and $\widetilde{H}$ (right) at $|P_3| = 1.25~\rm{GeV}$. For better presentation, we only show $x\geq0$. Results are given in the $\MSb$ scheme at 2 GeV.} 
    \label{fig:3D_plots}
\end{figure}

\newpage
\section{Summary and Future prospects}

This work builds upon recent advancements that enable the extraction of the $x$ dependence of GPDs from lattice QCD matrix elements calculated in any kinematic frame~\cite{Bhattacharya:2022aob}.
The main motivation is to efficiently compute GPDs across a range of $-t$ values with efficient use of computational resources, a task that proves challenging in the symmetric kinematic frame.
Our approach hinges on the decomposition of matrix elements into Lorentz invariant amplitudes, which can then be related to the quasi-GPDs.
In this work, we concentrate on the axial-vector case for the proton, providing a detailed framework to extract the helicity GPDs, $\widetilde{H}$ and $\widetilde{E}$.
To illustrate our methodology, we present a proof-of-concept calculation, where we obtain the matrix element in the standard symmetric frame, as well as an asymmetric frame, in which the momentum transfer is assigned to the initial state of the proton; the parameters are chosen to lead to a very similar value of $-t$.
The calculation is performed at zero skewness, which has the limitation that the $\widetilde{E}$ cannot be obtained directly from the lattice data.
Our analysis involves a comparison of lattice data for the Lorentz invariant amplitudes, $\widetilde{A}_i$, which confirms the theoretical expectations that the amplitudes are frame independent.
Such a finding paves the way for a complete asymmetric frame calculation to obtain the light-cone GPD $\widetilde{H}$ at multiple values of $-t$.
In our work, we employ two definitions for the quasi-GPDs, namely the standard definition of $\gamma_3\gamma_5$, as well as a Lorentz invariant definition that is based on the same
functional form in terms of the $\widetilde{A}_i$ as the light-cone GPDs.
Our observations suggest that both definitions yield comparable numerical results.
However, it is important to highlight that the constructed Lorentz invariant definition receives contributions from finite mixing of different operators under renormalization due to chiral symmetry breaking~\cite{Constantinou:2017sej}.

Our final results for the GPDs are presented in the $\MSb$ scheme at a scale of 2 GeV and are summarized in Figs.~\ref{fig:t_dependence_GPD} - \ref{fig:3D_plots} for various $-t$ values covering the range $[0.17,\,2.77]$ GeV$^2$. 
While we have achieved a robust signal, reducing statistical noise further proves to be a formidable challenge for off-forward matrix elements. 
Considering the fact that systematic uncertainties are still unquantified for GPDs, aiming for high statistical accuracy falls outside the scope of this work.
In the future, we anticipate a deeper exploration of systematic uncertainties, including the effects of eliminating excited states, as well as the impact of volume and discretization effects. 
We also aim to investigate the dependence on the pion mass and other systematic factors tied to the quasi-distribution approach, such as finite momentum boost and limitations of the one-loop formalism. 
There is also the potential for parameterizing the $-t$ dependence and exploring the impact-parameter space across a broad range of $-t$ values. 
Another avenue is introducing nonzero skewness, which would facilitate the direct extraction of $\widetilde{\mathcal{E}}$ from lattice data and allow for extrapolation to $\xi=0$.

In summary, this study underscores the significant potential for advancing lattice QCD calculations with regard to generalized parton distributions. 
These advances promise precision calculations that will contribute to developing a framework of global analysis of both current and forthcoming experimental data, in which incorporating lattice data is possible.

\begin{acknowledgements}
The Authors are grateful to Aurora Scapellato for her contributions to the initial stages of this project.
S.~B., S.~M., and P.~P. are supported by the U.S. Department of Energy, Office of Science, Office of Nuclear Physics through Contract No. DE-SC0012704. S.~B. is also supported by Laboratory Directed Research and Development (LDRD) funds from Brookhaven Science Associates. 
K.~C.\ is supported by the National Science Centre (Poland) grants SONATA BIS no.\ 2016/22/E/ST2/00013 and OPUS no.\ 2021/43/B/ST2/00497. M.~C., J. D., J.~M. and A.~S. acknowledge financial support by the U.S. Department of Energy, Office of Nuclear Physics, Early Career Award under Grant No.\ DE-SC0020405.
 J. D. also received support by the U.S. Department of Energy, Office of Science, Office of Nuclear Physics, within the framework of the TMD Topical Collaboration. 
The work of A.~M. has been supported by the National Science Foundation under grant number PHY-2110472, and also by the U.S. Department of Energy, Office of Science, Office of Nuclear Physics, within the framework of the TMD Topical Collaboration. 
F.~S.\ was funded by the NSFC and the Deutsche Forschungsgemeinschaft (DFG, German Research Foundation) through the funds provided to the Sino-German Collaborative Research Center TRR110 “Symmetries and the Emergence of Structure in QCD” (NSFC Grant No. 12070131001, DFG Project-ID 196253076 - TRR 110). 
Y.~Z. and X.~G. received support by the U.S. Department of Energy, Office of Science, Office of Nuclear Physics through Contract No.~DE-AC02-06CH11357. X.~G. was also partially supported by the U.S. Department of Energy, Office of Science, Office of Nuclear Physics within the frameworks of Scientific Discovery through Advanced Computing (SciDAC) award \textit{Fundamental Nuclear Physics at the Exascale and Beyond} and the Quark-Gluon Tomography (QGT) Topical Collaboration, under contract no.~DE-SC0023646.
Computations for this work were carried out in part on facilities of the USQCD Collaboration, which are funded by the Office of Science of the U.S. Department of Energy. 
This research used resources of the National Energy Research Scientific Computing Center, a DOE Office of Science User Facility supported by the Office of Science of the U.S. Department of Energy under Contract No. DE-AC02-05CH11231 using NERSC award NP-ERCAP0022961.
This research was supported in part by PLGrid Infrastructure (Prometheus supercomputer at AGH Cyfronet in Cracow).
Computations were also partially performed at the Poznan Supercomputing and Networking Center (Eagle supercomputer), the Interdisciplinary Centre for Mathematical and Computational Modelling of the Warsaw University (Okeanos supercomputer), and at the Academic Computer Centre in Gda\'nsk (Tryton supercomputer). The gauge configurations have been generated by the Extended Twisted Mass Collaboration on the KNL (A2) Partition of Marconi at CINECA, through the Prace project Pra13\_3304 ``SIMPHYS".
Inversions were performed using the DD-$\alpha$AMG solver~\cite{Frommer:2013fsa} with twisted mass support~\cite{Alexandrou:2016izb}. 
\end{acknowledgements}


\appendix

\section{Alternative parameterization for the axial-vector matrix element}
\label{s:another_basis}
In this appendix, we present a representation of the matrix element in Eq.~\eqref{helicity_para} using an alternative basis:
\begin{align}
\widetilde{F}^{' \mu} (z, P, \Delta) & = \bar{u}(p_f,\lambda') \bigg [ \dfrac{i \epsilon^{\mu P z \Delta}}{m} \widetilde{A}'_1 + \dfrac{i \sigma^{\mu P} \gamma_5}{m} \widetilde{A}'_2 \nonumber \\[0.2cm]
& + i \sigma^{P z} \gamma_5 \bigg ( \dfrac{P^\mu}{m} \widetilde{A}'_3 + m z^\mu \widetilde{A}'_4 + \dfrac{\Delta^\mu}{m} \widetilde{A}'_5 \bigg ) + \dfrac{i \sigma^{P\Delta} \gamma_5}{m^2} \bigg ( \dfrac{P^\mu}{m} \widetilde{A}'_6 + m z^\mu \widetilde{A}'_7 + \dfrac{\Delta^\mu}{m} \widetilde{A}'_8 \bigg )\bigg ] u(p_i, \lambda) \, ,
\label{helicity_para_basis2}
\end{align}
where $\sigma^{\mu P}=\sigma^{\mu \alpha} P_\alpha$, $\sigma^{Pz}=\sigma^{\alpha \beta} P_\alpha z_\beta$ and so on. In this basis, the connections between the Lorentz-invariant (LI) quasi-GPDs (bearing the same functional form as light-cone GPDs) and the amplitudes are as follows:
\begin{align}
\widetilde{\mathcal{H}}' (z \cdot P, z \cdot \Delta, \Delta^2, z^2) & = \widetilde{A}'_2 - (P \cdot z) \widetilde{A}'_3 - (\Delta \cdot z) \widetilde{A}'_5 \, , \label{H_LC_basis2} \\[0.2cm]
\widetilde{\mathcal{E}}' (z \cdot P, z \cdot \Delta, \Delta^2,z^2) & = - \widetilde{A}'_2 + (P \cdot z) \widetilde{A}'_3 + (\Delta \cdot z) \widetilde{A}'_5 -  \dfrac{4P^2}{m^2} \dfrac{ P \cdot z}{\Delta \cdot z} \widetilde{A}'_6 - \dfrac{4P^2}{m^2} \widetilde{A}'_8 \, .
\label{E_LC_basis2}
\end{align}
It can be verified that the amplitudes $\widetilde{A}_i$, defined through Eq.~\eqref{helicity_para} in Section~\ref{s:para}, and the $\widetilde{A}'_i$ are related as follows:
\begin{align}
\widetilde{A}_1 & = \widetilde{A}'_1 \, ,\nonumber \\[0.2cm]
\widetilde{A}_2 & = \widetilde{A}'_2 \, ,\nonumber \\[0.2cm]
\widetilde{A}_3 & = \dfrac{\Delta \cdot z}{2} \widetilde{A}'_3 - \dfrac{2P^2}{m^2} \widetilde{A}'_6 \, ,\nonumber \\[0.2cm]
\widetilde{A}_4 & = \dfrac{\Delta \cdot z}{2} \widetilde{A}'_4 - \dfrac{2P^2}{m^2} \widetilde{A}'_7 \, ,\nonumber \\[0.2cm]
\widetilde{A}_5 & = - \dfrac{\widetilde{A}'_2}{2} + \dfrac{\Delta \cdot z}{2} \widetilde{A}'_5 - \dfrac{2P^2}{m^2} \widetilde{A}'_8 \, ,\nonumber \\[0.2cm]
\widetilde{A}_6 & = - \widetilde{A}'_3 \, ,\nonumber \\[0.2cm]
\widetilde{A}_7 & = -  \widetilde{A}'_4 \, ,\nonumber \\[0.2cm]
\widetilde{A}_8 & = -  \widetilde{A}'_5 \, .
\label{e:amps_alt_basis}
\end{align}
We would like to emphasize that these relations play a pivotal role in demonstrating the equivalence of the LI quasi-GPD results in both bases. This finding holds significant importance as it ensures the uniqueness of the construction of LI definitions for quasi-GPDs, thereby securing consistent numerical values regardless of the basis. Regarding the light-cone GPDs, their uniqueness arises from their distinct relationship to matrix elements. However, one can further strengthen this assertion by explicitly verifying it using the aforementioned amplitude expressions, Eq.~(\ref{e:amps_alt_basis}).

On the other hand, the connections between the quasi-GPDs utilizing the operator $\gamma^3 \gamma_5$ and the amplitudes can be described as follows: 
\begin{align}
\widetilde{\mathcal{H}}'_3 (z, P, \Delta) & = \widetilde{A}'_2 + z^3 P^3 \widetilde{A}'_3 + m^2 (z^3)^2 \widetilde{A}'_4 + z^3 \Delta^3 \widetilde{A}'_5 \, ,\\[0.2cm]
\widetilde{\mathcal{E}}'_3 (z, P, \Delta) & = - \widetilde{A}'_2 - z^3 P^3 \widetilde{A}'_3 - m^2 (z^3)^2 \widetilde{A}'_4 - z^3 \Delta^3 \widetilde{A}'_5 - \dfrac{4P^2}{m^2} \dfrac{P^3}{\Delta^3} \widetilde{A}'_6 - 4 P^2 \dfrac{z^3}{\Delta^3} \widetilde{A}'_7 - \dfrac{4P^2}{m^2} \widetilde{A}'_8 \, .
\label{amp_qgpds_basis2}
\end{align}
Once again, by using Eq.~(\ref{e:amps_alt_basis}), one can affirm the equivalence of the results for the quasi-GPDs utilizing the operator $\gamma^3\gamma_5$ in both of the aforementioned bases. As also discussed in Section~\ref{s:para}, these expressions can be reformulated in a Lorentz-invariant manner, presenting additional contenders for LI definitions of quasi-GPDs alongside the definitions in Eqs.~(\ref{H_LC_basis2})-(\ref{E_LC_basis2}). 

Note that while here we have shown the uniqueness of the definition of the LI quasi-GPDs for just one alternative basis, our procedure of constructing the LI quasi-GPDs should provide the same numerical result for \textit{any} basis.  The only requirement one has to satisfy is that the (new) basis vectors are such that the relations between the (new) amplitudes and the ones defined in Eq.~\eqref{helicity_para} do not contain factors of $z^2$. Otherwise there might be the possibility of numerical disparities because our approach of constructing LI definitions involve the elimination of explicit factors of $z^2$. Instead, we implicitly capture all the $z^2$-dependence within the amplitudes themselves.

\section{Symmetry property of the $\widetilde{A}_i$ amplitudes}
\label{s:symmetries}
Here, we provide a summary of the symmetry properties of the amplitudes implied by Hermiticity and the time-reversal transformation. To derive these properties, we closely follow the steps outlined in our previous work~\cite{Bhattacharya:2022aob}.

\textbf{\textit{Symmetry of the $\widetilde{A}_i$ following from Hermiticity}:}
\begin{align}
- \widetilde{A}^*_1 (- z \cdot P, z \cdot \Delta, \Delta^2, z^2) & = \widetilde{A}_1 (z \cdot P, z \cdot \Delta, \Delta^2, z^2) \, , \nonumber \\[0.2cm]
 \widetilde{A}^*_2 (- z \cdot P, z \cdot \Delta, \Delta^2, z^2) & = \widetilde{A}_2 (z \cdot P, z \cdot \Delta, \Delta^2, z^2) \, , \nonumber \\[0.2cm]
- \widetilde{A}^*_3 (- z \cdot P, z \cdot \Delta, \Delta^2, z^2) & = \widetilde{A}_3 (z \cdot P, z \cdot \Delta, \Delta^2, z^2) \, , \nonumber \\[0.2cm]
\widetilde{A}^*_4 (- z \cdot P, z \cdot \Delta, \Delta^2, z^2) & = \widetilde{A}_4 (z \cdot P, z \cdot \Delta, \Delta^2, z^2) \, , \nonumber \\[0.2cm]
\widetilde{A}^*_5 (- z \cdot P, z \cdot \Delta, \Delta^2, z^2) & = \widetilde{A}_5 (z \cdot P, z \cdot \Delta, \Delta^2, z^2) \, , \nonumber \\[0.2cm]
- \widetilde{A}^*_6 (- z \cdot P, z \cdot \Delta, \Delta^2, z^2) & = \widetilde{A}_6 (z \cdot P, z \cdot \Delta, \Delta^2, z^2) \, , \nonumber \\[0.2cm]
\widetilde{A}^*_7 (- z \cdot P, z \cdot \Delta, \Delta^2, z^2) & = \widetilde{A}_7 (z \cdot P, z \cdot \Delta, \Delta^2, z^2) \, , \nonumber \\[0.2cm]
\widetilde{A}^*_8 (- z \cdot P, z \cdot \Delta, \Delta^2, z^2) & = \widetilde{A}_8 (z \cdot P, z \cdot \Delta, \Delta^2, z^2) \, .
\label{e:hermicity_trans}
\end{align}
The relations in~\eqref{e:hermicity_trans} indicate, for instance,  the symmetry of the amplitudes under the transformation $P^3 \rightarrow -P^3$ while keeping $z$ fixed.

\textbf{\textit{Symmetry of the $\widetilde{A}_i$ following from time-reversal}:}
\begin{align}
- \widetilde{A}^*_1 (- \bar{z} \cdot \bar{P}, -\bar{z} \cdot \bar{\Delta}, \bar{\Delta}^2, \bar{z}^2) & = \widetilde{A}_1 (z \cdot P, z \cdot \Delta, \Delta^2, z^2) \, , \nonumber \\[0.2cm]
\widetilde{A}^*_2 (- \bar{z} \cdot \bar{P}, -\bar{z} \cdot \bar{\Delta}, \bar{\Delta}^2, \bar{z}^2) & = \widetilde{A}_2 (z \cdot P, z \cdot \Delta, \Delta^2, z^2) \, , \nonumber \\[0.2cm]
\widetilde{A}^*_3 (- \bar{z} \cdot \bar{P}, -\bar{z} \cdot \bar{\Delta}, \bar{\Delta}^2, \bar{z}^2) & = \widetilde{A}_3 (z \cdot P, z \cdot \Delta, \Delta^2, z^2) \, , \nonumber \\[0.2cm]
- \widetilde{A}^*_4 (- \bar{z} \cdot \bar{P}, -\bar{z} \cdot \bar{\Delta}, \bar{\Delta}^2, \bar{z}^2) & = \widetilde{A}_4 (z \cdot P, z \cdot \Delta, \Delta^2, z^2) \, , \nonumber \\[0.2cm]
\widetilde{A}^*_5 (- \bar{z} \cdot \bar{P}, -\bar{z} \cdot \bar{\Delta}, \bar{\Delta}^2, \bar{z}^2) & = \widetilde{A}_5 (z \cdot P, z \cdot \Delta, \Delta^2, z^2) \, , \nonumber \\[0.2cm]
- \widetilde{A}^*_6 (- \bar{z} \cdot \bar{P}, -\bar{z} \cdot \bar{\Delta}, \bar{\Delta}^2, \bar{z}^2) & = \widetilde{A}_6 (z \cdot P, z \cdot \Delta, \Delta^2, z^2) \, , \nonumber \\[0.2cm]
\widetilde{A}^*_7 (- \bar{z} \cdot \bar{P}, -\bar{z} \cdot \bar{\Delta}, \bar{\Delta}^2, \bar{z}^2) & = \widetilde{A}_7 (z \cdot P, z \cdot \Delta, \Delta^2, z^2) \, , \nonumber \\[0.2cm]
- \widetilde{A}^*_8 (- \bar{z} \cdot \bar{P}, -\bar{z} \cdot \bar{\Delta}, \bar{\Delta}^2, \bar{z}^2) & = \widetilde{A}_8 (z \cdot P, z \cdot \Delta, \Delta^2, z^2) \, ,
\label{e:time_trans}
\end{align}
where $\bar{z} = (z^0, -\vec{z})$ and so on.

{\textbf{\textit{Symmetry of the $\widetilde{A}_i$ following from Hermiticity and time-reversal}}}: One can combine the relations in~\eqref{e:hermicity_trans} and~\eqref{e:time_trans} to find the following constraints:

\begin{align}
\widetilde{A}_1 (\bar{z} \cdot \bar{P}, -\bar{z} \cdot \bar{\Delta}, \bar{\Delta}^2, \bar{z}^2) & = \widetilde{A}_1 (z \cdot P, z \cdot \Delta, \Delta^2, z^2) \, , \nonumber \\[0.2cm]
\widetilde{A}_2 (\bar{z} \cdot \bar{P}, -\bar{z} \cdot \bar{\Delta}, \bar{\Delta}^2, \bar{z}^2) & = \widetilde{A}_2 (z \cdot P, z \cdot \Delta, \Delta^2, z^2) \, , \nonumber \\[0.2cm]
- \widetilde{A}_3 (\bar{z} \cdot \bar{P}, -\bar{z} \cdot \bar{\Delta}, \bar{\Delta}^2, \bar{z}^2) & = \widetilde{A}_3 (z \cdot P, z \cdot \Delta, \Delta^2, z^2) \, , \nonumber \\[0.2cm]
- \widetilde{A}_4 (\bar{z} \cdot \bar{P}, -\bar{z} \cdot \bar{\Delta}, \bar{\Delta}^2, \bar{z}^2) & = \widetilde{A}_4 (z \cdot P, z \cdot \Delta, \Delta^2, z^2) \, , \nonumber \\[0.2cm]
\widetilde{A}_5 (\bar{z} \cdot \bar{P}, -\bar{z} \cdot \bar{\Delta}, \bar{\Delta}^2, \bar{z}^2) & = \widetilde{A}_5 (z \cdot P, z \cdot \Delta, \Delta^2, z^2) \, , \nonumber \\[0.2cm]
\widetilde{A}_6 (\bar{z} \cdot \bar{P}, -\bar{z} \cdot \bar{\Delta}, \bar{\Delta}^2, \bar{z}^2) & = \widetilde{A}_6 (z \cdot P, z \cdot \Delta, \Delta^2, z^2) \, , \nonumber \\[0.2cm]
\widetilde{A}_7 (\bar{z} \cdot \bar{P}, -\bar{z} \cdot \bar{\Delta}, \bar{\Delta}^2, \bar{z}^2) & = \widetilde{A}_7 (z \cdot P, z \cdot \Delta, \Delta^2, z^2) \, , \nonumber \\[0.2cm]
- \widetilde{A}_8 (\bar{z} \cdot \bar{P}, -\bar{z} \cdot \bar{\Delta}, \bar{\Delta}^2, \bar{z}^2) & = \widetilde{A}_8 (z \cdot P, z \cdot \Delta, \Delta^2, z^2) \, .
\label{e:HTR_cons}
\end{align}
The relations in~\eqref{e:HTR_cons} immediately provide the symmetry behavior of the amplitudes under the transformation $\xi \rightarrow -\xi$.

\section{Consistency with the local case $z = 0$}
\label{s:local}
It is important to verify the consistency of our decomposition with the local axial-vector current. We remind the reader that the local axial-vector operator, which defines the axial-vector ($g_A$) and pseudo-scalar ($g_P$) form factors, is given by
\begin{align}
\langle p_f,\lambda '| \bar{\psi}(0)\gamma^{\mu} \gamma_5 \psi(0) |p_i,\lambda \rangle &= \bar{u} (p_f,\lambda ') \bigg [ \gamma^\mu \gamma_5 g_A (\Delta^2) + \dfrac{\Delta^\mu \gamma_5}{2m} g_P (\Delta^2) \bigg ] u (p_i,\lambda ) \, .
\label{e:ga_gp}
\end{align}
These two form factors are real functions. Conversely, in the local case $z=0$, our decomposition simplifies to the following expression:
\begin{align}
\widetilde{F}^{\mu} \big |_{z=0} & = \bar{u}(p_f,\lambda') \bigg [ \gamma^\mu \gamma_5 \widetilde{A}_2 + \gamma_5 \dfrac{P^{\mu}}{m} \widetilde{A}_3 + \gamma_5 \dfrac{\Delta^\mu}{m} \widetilde{A}_5 \bigg ] u(p_i, \lambda) \,.
\label{e:local_limit}
\end{align}
Now, recall that, generally, the $\widetilde{A}_i$ are complex amplitudes. However, in order to maintain consistency with the local axial-vector operator, we need to demonstrate that the surviving $\widetilde{A}_i$ coefficients are real. In other words, it is essential to note that only either the real part or the imaginary part of the amplitudes in Eq.~\eqref{e:local_limit} can be nonzero, but not both simultaneously. Hermiticity leads to the following condition (see Appendix~\ref{s:symmetries}):
\begin{align}
\widetilde{A}^*_2 & = \widetilde{A}_2 \, , \qquad
- \widetilde{A}^*_3  = \widetilde{A}_3 \, , \qquad
\widetilde{A}^*_5  = \widetilde{A}_5 \, .
\end{align}
Subsequently, these constraints on the coefficients $\widetilde{A}_i$ give rise to the following implications:
\begin{align}
\textrm{Im.} (\widetilde{A}_2)  = 0 \, , \qquad \textrm{Re.} (\widetilde{A}_3)  = 0 \, , \qquad \textrm{Im.} (\widetilde{A}_5)  = 0 \, .
\label{e:c1}
\end{align}
Unfortunately, this approach is not entirely conclusive as it still leaves us with three coefficients, $\widetilde{A}_i$. In order to establish that only two coefficients remain, we need to investigate if the time-reversal transformation imposes any additional constraints on the $\widetilde{A}_i$'s. It is worth recalling that time-reversal leads to the following transformation (see Appendix~\ref{s:symmetries}):
\begin{align}
\widetilde{A}^*_2  & = \widetilde{A}_2  \, , \qquad
\widetilde{A}^*_3   = \widetilde{A}_3 \, , \qquad
\widetilde{A}^*_5   = \widetilde{A}_5  \, .
\end{align}
These constraints on the $\widetilde{A}_i$'s result in the following implications:
\begin{align}
\textrm{Im.} (\widetilde{A}_2)  = 0 \, , \qquad \textrm{Im.} (\widetilde{A}_3)  = 0 \, , \qquad \textrm{Im.} (\widetilde{A}_5) = 0 \, .
\label{e:c2}
\end{align}
By combining equations (\ref{e:c1}) and (\ref{e:c2}), we can draw the following conclusion:
\begin{align}
\textrm{Re.} (\widetilde{A}_3) & = 0 \, , \qquad \textrm{Im.} (\widetilde{A}_3) = 0 \nonumber \\[0.2cm]
\therefore \widetilde{A}_3 & = 0 \, .
\end{align}
Therefore, equations (\ref{e:c1}) and (\ref{e:c2}) indicate that the sole contribution at $z=0$ arises from:
\begin{align}
\textrm{Re.} (\widetilde{A}_2) \neq  0 \, , \qquad \textrm{Re.} (\widetilde{A}_5)\neq 0 \, .
\end{align}
Thus, our decomposition demonstrates consistency with the local axial-vector current:
\begin{align}
\widetilde{F}^{\mu} \big |_{z=0} & = \bar{u}(p_f,\lambda') \bigg [ \gamma^\mu \gamma_5 \widetilde{A}_2 + \gamma_5 \dfrac{\Delta^\mu}{m} \widetilde{A}_5 \bigg ] u(p_i, \lambda) \,.
\end{align}
By comparing this expression to Eq.~(\ref{e:ga_gp}), we can deduce that $g_A$ corresponds to $\widetilde{A}_2$ and $g_P$ corresponds to $\widetilde{A}_5$.

\bibliography{references.bib}

\end{document}